%% file: ms.tex
\documentclass[preprint]{aastex}

\begin{document}
%\input amssym.def \relax
%\newsymbol\lesssim 132E
%\newsymbol\gtrsim 1326

\title{Neutron Capture Elements in {\it s}-Process-Rich, Very Metal-Poor Stars}

\author{Wako Aoki\altaffilmark{1}, Sean G. Ryan\altaffilmark{2}, John E.
Norris\altaffilmark{3}, Timothy C. Beers\altaffilmark{4}, Hiroyasu
Ando\altaffilmark{1,5}, Nobuyuki Iwamoto\altaffilmark{1,6}, Toshitaka
Kajino\altaffilmark{1,5}, Grant J. Mathews\altaffilmark{7}, Masayuki Y.
Fujimoto\altaffilmark{8}}
\altaffiltext{1}{National Astronomical Observatory, Mitaka, Tokyo, 181-8588
Japan; email: aoki.wako@nao.ac.jp, ando@optik.mtk.nao.ac.jp,
iwamoto@th.nao.ac.jp, kajino@nao.ac.jp}
\altaffiltext{2}{Department of Physics and Astronomy, The Open University,
Walton Hall, Milton Keynes, MK7 6AA, UK; email: s.g.ryan@open.ac.uk}
\altaffiltext{3}{Research School of Astronomy and Astrophysics, The
Australian National University, Private Bag, Weston Creek Post Office,
Canberra, ACT 2611, Australia; email: jen@mso.anu.edu.au}
\altaffiltext{4}{Department of Physics and Astronomy, Michigan State
University, East Lansing, MI 48824-1116, USA; email: beers@pa.msu.edu}
\altaffiltext{5}{Department of Astronomy, Graduate School of Science,
University of Tokyo, Bunkyo-ku, Tokyo 113-0033, Japan}
\altaffiltext{6}{Center for Nuclear Study, University of Tokyo, Wako,
Saitama 351-0198, Japan}
\altaffiltext{7}{Department of Physics and Center for Astrophysics,
University of Notre Dame, Notre Dame, IN 46556, USA; email: gmathews@nd.edu}
\altaffiltext{8}{Department of Physics, Hokkaido University,
Sapporo, Hokkaido 060-0810, Japan; email: fujimoto@astro1.sci.hokudai.ac.jp}

\begin{abstract} 

We report abundance estimates for neutron-capture elements, including lead
(Pb), and nucleosynthesis models for their origin, in two carbon-rich, very
metal-poor stars, LP~625-44 and LP~706-7. These stars are subgiants whose
surface abundances are likely to have been strongly affected by mass transfer
from companion AGB stars that have since evolved to white dwarfs. The
detections of Pb, which forms the final abundance peak of the $s$-process,
enable a comparison of the abundance patterns from Sr ($Z=38$) to Pb ($Z=82$)
with predictions of AGB models. The derived chemical compositions provide
strong constraints on the AGB stellar models, as well as on $s$-process
nucleosynthesis at low metallicity. The present paper reports details of the
abundance analysis for 16 neutron-capture elements in LP~625-44, including the
effects of hyperfine splitting and isotope shifts of spectral lines for some
elements.  A Pb abundance is also derived for LP~706-7 by a re-analysis of a
previously observed spectrum.  We investigate the characteristics of the
nucleosynthesis pathway that produces the abundance ratios of these objects
using a parametric model of the {\it s}-process without adopting any specific
stellar model. The neutron exposure $\tau$ is estimated to be about
0.7mb$^{-1}$, significantly larger than that which best fits solar-system
material, but consistent with the values predicted by models of moderately
metal-poor AGB stars. This value is strictly limited by the Pb abundance, in
addition to those of Sr and Ba. We also find that the observed abundance
pattern can be explained by a few recurrent neutron exposures, and that the
overlap of the material that is processed in two subsequent exposures is small
(the overlap factor $r \sim 0.1)$.

\end{abstract}


\section{Introduction}\label{sec:intro}

Many efforts have been made to explain the solar-system abundances of elements
associated with the slow neutron-capture process ({\it s}-process).  One common
approach is the so-called {\it classical model}, which assumes an exponential
distribution of neutron exposures \citep{kappeler89}.  Use of this approach 
led to the conclusion that three distinct distributions of neutron exposures
are required to represent solar-system {\it s}-process abundances. One is
referred to as the {\it main component}, thought to be responsible
for most of the isotopes of {\it
s}-process origin with $90<A<204$ ($A$ indicates the mass number). Since the
elements with $A<90$ cannot be explained by the main component alone, another
distribution, with lower neutron exposure, was introduced (the so-called {\it
weak component}). The sites of the main and weak {\it s}-processes are believed
to be thermally-pulsing asymptotic giant branch (AGB) stars and helium core
burning massive stars, respectively. The solar-system abundances of the
heaviest nuclei, with $204\leq A \leq 209$, most of which are isotopes of Pb,
cannot be reproduced by these two components, so the third so-called {\it
strong component} with very high neutron exposure was introduced.

While this simple approach has been somewhat successful explaining the
solar-system {\it s}-process abundances, detailed models of
nucleosynthesis in thermally pulsing AGB stars have also been studied
in attempts to confront the data with specific predictions from the
likely production site.  Recent modeling of AGB stars by
\citet{straniero95} showed that neutron capture mainly occurs, not in
the convective He shell {\it during} a thermal pulse, but in the
radiative state {\it between} two given pulses. In this model the
density distribution of the $^{13}$C-rich layer (referred to as the
$^{13}$C pocket), which provides neutrons for the {\it s}-process, is
taken as a free parameter.  Since, in this case, the distribution of
neutron exposures cannot be approximated by an exponential, the yields
of neutron-capture elements have been systematically calculated based
on the stellar models by \citet{gallino98} and \citet{arlandini99},
who succeeded in reproducing at least the main component of the
solar-system {\it s}-process elements.

In the Torino AGB models mentioned above, the $^{13}$C pocket is generated
artificially and described parametrically, assuming that some (poorly
characterized) mixing of the overlying H-rich layers down into the He-rich
intershell region occurs.  A different approach to this problem is taken by the
Geneva group, who model the mixing process via a diffusion mechanism
\citep[e.g., ][]{goriely00}. Whichever approach is taken --- parameterization of
the $^{13}$C pocket by the Torino group or parameterization of mixing by the
Geneva group --- fundamental uncertainties currently exist in the models. It is
in the spirit of improving our understanding of these stars, rather than
confronting or endorsing any particular approach to modeling, that we present
the following analysis.

The calculation of {\it s}-process yields was extended to lower metallicity by
\citet{gallino98}, and a systematic investigation was performed by
\citet{busso99}.  Since the difference in metallicity of the model affects the
ratio of neutrons to seed nuclei, the distribution of {\it s}-process elements
produced in AGB stars should be sensitive to the metallicity as well.  While
the abundance of seed nuclei (most of which are iron) is proportional to the
metallicity, the production of $^{13}$C (the main neutron source) is expected
to be metallicity independent. Consequently, higher abundances of heavier elements
are expected in the yields of AGB stars with lower metallicity. The calculations
of \citet{gallino98} and \citet{busso99} indeed predict higher abundance ratios
of heavy to light {\it s}-process elements, and very high Pb and Bi abundances,
in the nucleosynthesis products of metal-poor AGB stars. Moreover, they
suggested that the origin of Pb, and hence the site of the strong component of
the {\it s}-process, should be attributed to these metal-poor AGB stars.

The nucleosynthesis of heavy elements in metal-poor AGB stars can be
investigated by abundance studies of carbon-rich and {\it s}-process-rich
objects, often referred to as CH stars, whose surface chemical composition is
considered to result from mass transfer from a now-extinct AGB companion. 
In one such star, LP~625-44, a Pb {\small I} line was detected by
\citet{aoki00} (hereafter Paper I) for the first time (in a CH star), and it
became possible to compare abundance ratios for elements from Sr to Pb with
model predictions.  LP~625-44 is an ideal object for this study.  One reason is
that it is very metal-poor ([Fe/H]$=-2.7$) and shows very high {\it s}-process
overabundances (e.g., [Ba/Fe]$=2.7$), so the abundances of heavy elements
almost purely represent the yields of the AGB donor (see Section
\ref{sec:disc1}). Another reason is that the variation of radial velocity, with
a period longer than 12 years (as found by our monitoring), strongly supports
the mass-transfer scenario.  The Pb abundance derived is, however, much lower
than the prediction by the standard model of \citet{busso99}. This result
provides a strong constraint on the nature of the $^{13}$C pocket,  which is a
parameter in their model \citep{ryan01}, and may even prompt re-consideration
of models of $s$-process nucleosynthesis in very metal-poor AGB stars.
Clearly, the study of elemental abundances in these objects is important for
investigation of the origin of Pb in the solar system.

In this paper, we report details of the abundance analysis of LP~625-44 that
was summarized in Paper {\small I}, and re-analyse an extended line list.  For
the analysis of many lines of neutron-capture elements, the effects of
hyperfine splitting and isotope shifts are taken into consideration. The line
data and these additional effects are described in Section \ref{sec:ana} and in
the Appendix. We also analyse another {\it s}-process rich, very metal-poor
star, LP~706-7, previously studied by \citet{norris97a}, and determine its Pb
abundance for the first time (Section {\ref{sec:ana}).  In section
\ref{sec:disc}, we discuss the characteristics of the nucleosynthesis which
produces the abundance ratios of these objects by a parametric model of the
{\it s}-process, without reference to any specific stellar model.

\section{Observation and Measurements}\label{sec:obs}

LP~625-44 and LP~706-7 were observed with the University College London coud\'e
\'echelle spectrograph (UCLES) and Tektronix 1024$\times$1024 CCD at the
Anglo-Australian Telescope. Our UV-blue spectra cover the wavelength region
from 3700 to 4700{\AA}. A red spectrum (5000-7800{\AA}) was also obtained for
LP~625-44. The observational and data reduction procedures have already been
reported in Paper I for LP~625-44, and in \citet{norris97a} for LP~706-7.
Details of the observations are summarized in Table 1. We note that the numbers
of photons obtained around the Pb {\small I} $\lambda$4057 are 8800 per 0.04\AA
\ pixel (S/N$\sim$150 per resolution element) and 3000 per 0.04\AA \ pixel
(S/N$\sim$80) for LP~625-44 and LP~706-7, respectively.

For LP~625-44, equivalent widths were measured for most elements by fitting
Gaussian profiles to the absorption lines. In Figure \ref{fig:compew} the
equivalent widths of \ion{Fe}{1}, measured in the present work, are compared
with those in \citet{norris96}, which were based on earlier spectra.  There is
no systematic difference between the two. We note that the S/N ratio in this
work is about twice that in \citet{norris96}.  The equivalent widths measured
for the lines of neutron-capture elements in LP~625-44 are listed in Table
\ref{tab:ew}.  For the elements Eu, Dy, Er, Tm, Hf and Pb, the abundances were
derived by spectrum synthesis. The equivalent widths given for these elements
(marked by daggers) in the table are the {\it synthesized} values that are
calculated for the abundance derived in our analysis.

\section{Abundance Analysis and Results}\label{sec:ana} 

\subsection{Stellar Atmosphere Parameters}\label{sec:atmos}

We carried out a standard abundance analysis based on the equivalent widths
and spectrum synthesis using model atmospheres in the ATLAS grid of
\citet{kurucz93a}. The stellar parameters have already been reported in Paper I
for LP~625-44. For the analysis of Pb {\small I} lines in LP~706-7, the
parameters determined by \citet{norris97a} were adopted. For convenience, we
summarize the stellar parameters (effective temperature:$T_{\rm eff}$, surface
gravity:$g$, micro-turbulent velocity:$v$ and iron abundance) in
Table~\ref{tab:para}.

The surface gravity of LP~706-7 \citep{norris97a} was based on the requirement
that \ion{Fe}{1} and \ion{Fe}{2} lines give identical abundances.  More
recently, a trigonometric parallax for this star has been published from the
{\it Hipparcos} mission \citep{esa97}, $\pi = 15.15 \pm 3.24$~mas.  Somewhat
surprisingly, this surface gravity indicates an absolute magnitude $M_{\rm V} =
8.0 \pm 0.4$, which is {\it sub-luminous} compared to both main sequence and
subgiant Population~II stars with $T_{\rm eff}$ = 6000~K. A subgiant of $M_{\rm
V}$ = 3.0 or 4.0 would have a parallax of only 1.5 or 2.4~mas.  Either the {\it
Hipparcos} measurement of this star is significantly in error, or the star is
far more bizarre than its CH-star status suggests. If the temperature estimate
(based on photometric colors) and the {\it Hipparcos} parallax were both
correct, we should be forced to infer a radius ten times smaller than for a
subgiant and four times smaller than for a main-sequence star, but the surface
gravity appears inconsistent with such a compact object (since $g \propto
M/R^2$). It seems most likely that the {\it Hipparcos} parallax is simply
incorrect, although an examination of the records (D. W. Evans, priv. comm.)
revealed no concerns.

An upper limit on the luminosity of LP~706-7 can be inferred via the
assumption that it is bound to the Galaxy, the local escape velocity
from which appears to be $v_{\rm esc} \sim 450-550$~km~s$^{-1}$
\citep{ryan91,allen91}. A luminosity as bright as $M_{\rm V}$ = 3.0
would imply a Galactic rest-frame velocity $v_{\rm RF}$ =
788~km~s$^{-1}$, considerably in excess of the escape value, whereas
$M_{\rm V}$ = 4.0 would imply $v_{\rm RF}$ = 464~km~s$^{-1}$,
consistent with the star being bound. This limit on the star's
luminosity supports the conclusion from its spectroscopic surface
gravity, and from the evolutionary state associated with its effective
temperature, that this object has not undergone first dredge-up. This
is particularly important for LP~706-7, because radial-velocity
variations that might be expected for a star with a white-dwarf
companion have {\it not} yet been detected \citep{norris97a}. In the
following we assume that LP~706-7 has been chemically enriched by a
similar process to that experienced by LP~625-44, but the differences
between these two stars (LP~706-7 being less evolved and exhibiting no
radial-velocity variations) should be kept in mind. (Some possible
alternative {\it s}-process sites to AGB stars are also discussed in
Section \ref{sec:disc1}.)

\subsection{Pb Abundance}\label{sec:pb}

In the UV-blue spectrum of the Sun, four Pb {\small I} lines have been
identified at 3639.5, 3683.4, 3739.9 and 4057.8\AA . Our spectra cover these last
two. \citet{youssef89} tried to analyse Pb in the solar photosphere using their
oscillator strengths of these lines. Despite the severe blending with lines of
other elements, a Pb abundance $\log \epsilon$ (Pb) $\sim 2.0$, consistent with
the meteoritic value, was derived from the two lines at 3739.9 and 4057.8\AA .
We adopted the line data determined by \citet{youssef89} in the present
analysis. The oscillator strengths of these lines agree well with the recent
result by \citet{biemont00}. 

In Figure \ref{fig:sp_pb}, the synthetic spectra around the Pb {\small
I} $\lambda$4057.8 for LP~625-44 and LP~706-7 are shown along with the
observed spectra. In this wavelength region the positions of CH lines
are identified at 4057.7\AA \ and 4058.2\AA, in addition to Mg {\small
I} 4057.5\AA. The Pb {\small I} $\lambda$4057 line is clearly
identified in LP~625-44, and also in LP~706-7, though it is much weaker in the
latter than in the former.

For comparison, we also show the spectra of HD~140283 and CS~22957-027 in the
figure. HD~140283 is a very metal-poor subgiant with similar physical parameters
to LP~625-44 and LP~706-7 ($T_{\rm eff}=5750$~K, $\log g=3.4$ and [Fe/H]
$=-$2.54, Ryan et al. 1996), but it exhibits no enhancement of neutron-capture
elements or of carbon. Since there is no distinct feature at 4057.8\AA \ in the
spectrum of HD~140283, the contamination of metal lines (arising from, e.g., 
$\alpha$-elements or iron-peak elements), whose abundances in HD~140283 are
comparable to those in our carbon-rich objects, is not large at this wavelength
in metal-poor subgiants like LP~625-44 and LP~706-7.  As a further check on
contamination due to CH and CN lines, which might be expected to be a problem
in carbon-rich stars, the observed and synthetic spectra of CS~22957-027 are
shown in the bottom of Figure~\ref{fig:sp_pb}. \citet{norris97b} showed that
this star is a very metal-poor giant ($T_{\rm eff}=4850$K, $\log g=1.9$ and
[Fe/H] $=-3.38$) with very large excesses of $^{12}$C, $^{13}$C and N, but no
excess of neutron-capture elements. The comparison of this spectrum with
those of LP~625-44 and LP~706-7 indicates that the absorption features at
4057.8\AA \ are {\it not} due to the presence of unrecognized CH and CN lines.

The solid lines in Figure~\ref{fig:sp_pb} are the synthetic spectra calculated
using our adopted model atmospheres. The Pb abundances assumed are
[Pb/Fe]=2.25, 2.55 and 2.85 for LP~625-44, [Pb/Fe]= 2.0, 2.3 and 2.6 for
LP~706-7, and [Pb/Fe]=0.0 for the other two stars.  The partition function of
Pb {\small I} derived by \citet{irwin81} was used in the analysis.  In the
calculation of the synthetic spectra, the effect of hyperfine splitting and
isotope shifts on the Pb {\small I} line is included, whereas the abundance
analysis of Paper~I used a single-line approximation.  (The data are given in
the Appendix, Table~A6).  These changes reduce the Pb abundance for LP~625-44
by 0.1~dex compared to the result in Paper I. The effect of these splittings is
much smaller for LP~706-7 because the Pb {\small I} line is weaker.  Table~A6
in the Appendix assumes the solar-system isotope ratio for Pb,
($^{204}$Pb:$^{206}$Pb:$^{207}$Pb:$^{208}$Pb = 0.015:0.236:0.226:0.523).  For
LP~625-44, we also tried the isotope ratio predicted for the $s$-process in AGB
stars \citep{arlandini99}, ($^{204}$Pb:$^{206}$Pb:$^{207}$Pb:$^{208}$Pb =
0.04:0.24:0.28:0.44), but found that the difference from the result derived
using the solar-system ratio is negligible. We adopted the Pb abundances
derived using the solar-system Pb isotope ratio, and list them in Table
\ref{tab:res}.

Another Pb {\small I} line covered by our blue spectra is Pb {\small
I} $\lambda$ 3739 ($\log gf=-0.12$ and $\chi=2.66$eV). However, no distinct
absorption feature appears at this wavelength in our spectra. The upper limit
on the Pb abundance ([Pb/Fe]$<+3.2$) derived for this line in LP~625-44 (Paper
I) is uninteresting. The same is true in LP~706-7.  No additional information
could be obtained from the Pb {\small I} $\lambda$7229 line \citep[$\log
gf=-1.61$ and $\chi=2.66$eV,][]{biemont00}, covered by the red spectrum of
LP~625-44, because of the weakness of this line.

\subsection{Other Neutron-Capture Elements}\label{sec:heavy}

Abundances of other neutron-capture elements besides Pb are also important to
understand nucleosynthesis at low metallicity. As shown by \citet{norris97a},
the abundances of neutron-capture elements in these two stars are basically
explained by the predictions of canonical {\it s}-process nucleosynthesis. We
previously reported neutron-capture abundances of LP~625-44 in Paper I, but
have improved them in the present effort by considering hyperfine splitting and
isotope shifts for as many lines as possible, and for some elements by studying
additional lines.  Here we summarize the details of the abundance analysis of
neutron-capture elements in LP~625-44.  We reviewed the spectrum of LP~706-7
used in \citet{norris97a} for additional elements, but added only Pb. For this
reason we used the abundances derived by \citet{norris97a} for neutron-capture
elements (other than Pb) for LP~706-7.

An extensive line list for many neutron-capture elements was compiled by
\citet{sneden96}, for their analysis of the {\it r}-process-enhanced star
CS~22892-052.  We supplemented their list with many new lines that were
detected in LP~625-44, because of its larger excess of neutron-capture
elements.  In Table~\ref{tab:ew}, we list the line data adopted in this work as
well as the equivalent widths measured in section~\ref{sec:obs}. Lines included
after Paper~I was completed are indicated by an asterisk.  Moreover, we include
the effects of hyperfine splitting and isotope shifts to the extent
possible for Ce, Pr, Nd and Sm, in addition to La and Eu, for which
\citet{norris97a} computed such effects in their subset of lines.  In
Table~\ref{tab:ew}, the abundances determined by the present analysis
{\it and} including the effect of splittings ($\delta(\log
\epsilon) = \log \epsilon_{\rm single line} - \log \epsilon_{\rm
HFS,IS}$) are given in the 5th and 6th columns, respectively. 

Below we describe the line data, and provide comments on the abundance analysis
for each element.  More detailed data on the hyperfine splitting and isotope
shifts are given in the Appendix.

%{\it Strontium, Yttrium, Zirconium}: 
\paragraph{Strontium, Yttrium, Zirconium:}
The line data listed by \citet{sneden96} were employed for the analysis of Sr
{\small II} and Y {\small II}. The {\it gf}-values measured by
\citet{biemont81}, also used by \citet{sneden96}, were adopted for Zr {\small
II}. While the two Sr {\small II} lines ($\lambda$4077.7 and $\lambda$4215.5)
are very strong, another line ($\lambda$4161.8) is quite weak. The abundance
derived from Sr {\small II} $\lambda$4161.8 is higher by 0.3 dex than those
derived from other strong lines, similar to the results reported by
\citet{sneden96}. We averaged the abundances derived from these three
lines.

%{\it Barium}: 
\paragraph{Barium:}
Since the Ba {\small II} resonance line at 4554\AA \ is extremely
strong (the equivalent width is 245m\AA), we excluded this line from
the abundance analysis. We used three Ba {\small II} lines in the red
spectrum, because the lines are not so strong as the resonance line,
and the effect of hyperfine splitting is small. \citet{norris97a} gave the
limit on the effect of hyperfine splitting in $\lambda$6496 line
as $<0.02$~dex for LP~625-44. We confirmed that the effect is also
smaller than 0.02~dex for the $\lambda$5853 and $\lambda$6141 lines using
the line data provided by \citet{mcwilliam98}. For these lines the
{\it gf}-values in \citet{sneden96} are adopted.

\paragraph{Lanthanum:} The {\it gf}-values listed by \citet{sneden96} were
used for five lines, while the values determined by \citet{bord96}
were adopted for others. The agreement of {\it gf}-values between
\citet{sneden96} and \citet{bord96} is fairly good. This element has
only one significant isotope ($^{139}$La). We included the effect of
hyperfine splitting for every line, though for most lines the splitting of the upper level is
unknown and assumed zero (see Appendix).  The $\delta$(log~$\epsilon$) values
of these lines must be regarded as uncertain, and are flagged by a ``:'' in
column~6 of Table~2.  Even with this coarse simplification, the effect of the
splitting is highly significant for several lines, $\simeq$1.5~dex for
$\lambda$3949, $\lambda$3988 and $\lambda$4238.  Although we did not include
the effect of hyperfine splitting in Paper I, the lines adopted there were quite weak and the
effect is not severe; the abundance from our new work is lower by only
0.04~dex, but now is based on 14 lines, rather than just 5 lines as in Paper
{\small I}.

\paragraph{Cerium:} We identified more than one hundred Ce {\small II} 
lines in the spectrum of LP~625-44 using the line list of \citet{cb62}. 
We selected 26 weak, unblended lines for the abundance analysis.  Since there
is no reliable source of oscillator strengths for most lines, we scaled the
{\it gf}-values of \citet{cb62} to be consistent to those determined by
\citet{gs94} ($\langle\log gf_{\rm GS94}-\log gf_{\rm CB62}\rangle$=+0.23).
Since all Ce isotopes have even-{\it N} nuclei, most being $^{140}$Ce and
$^{142}$Ce, there is no hyperfine splitting. We have approximated the isotope
shifts for the two significant isotopes (see Appendix), but find the effect on
abundances is quite small ($\leq$0.01~dex) in the weak lines selected in the
present analysis. Therefore the effect was neglected. The limits on $\delta
(\log \epsilon)$ shown in Table~\ref{tab:ew} are nevertheless flagged as
uncertain because of the approximate nature of the isotope shift calculation.

\paragraph{Praseodymium:} We adopted the {\it gf}-values measured by
\citet{goly91} for 6 lines. For other lines, \citet{cb62}
values scaled to \citet{goly91} are used ($\langle\log gf_{\rm Goly91}-\log
gf_{\rm CB62}\rangle$=+0.60).  Since there is only one stable isotope
$^{141}$Pr, there is no isotope shift, but hyperfine splitting for this odd-{\it N} nucleus is sometimes
significant, giving $\delta$(log~$\epsilon$) values up to 0.64.

\paragraph{Neodymium:} Following \citet{sneden96}, we adopted the oscillator
strength determined by \citet{maier77} and corrected the values by
\citet{ward85}. We re-selected the lines for which reliable {\it gf}-values
exist. Since we have no information on hyperfine splitting, we included only
the isotope shift, using the
solar-system isotope ratios.  As odd-{\it N} nuclei account for only 20\%\ of
the solar-system Nd isotopes, hyperfine effects should be small.  The maximum
effect computed, 0.15~dex, appears for the $\lambda$4061 and $\lambda$4462
lines.

\paragraph{Samarium:} We adopted the {\it gf}-values provided by
\citet{biemont89} for 6 lines and the \citet{cb62} values scaled to
\citet{biemont89} for the other 17 lines ($\langle\log gf_{\rm B89}-\log
gf_{\rm CB62}\rangle$=0.49).  There are seven main isotopes for Sm. While we
neglected hyperfine splitting due to insufficient information, we included the
isotope shifts, and assumed
solar-system isotope ratios.  As given in Table~\ref{tab:ew}, the effect of the
isotopic shift is small ($\leq$0.08); one of the reasons is that the lines selected in our
analysis are quite weak.  Therefore this effect should be negligible even if
the isotope ratios are different from the solar-system values.

\paragraph{Europium:} We adopted the $gf$-values determined by
\citet{biemont82}. The $\lambda$4205.5 line, which is frequently used for
abundance analysis, was excluded because of severe blending with CH molecular
lines. The $\lambda 3930$ line was detected, but is strongly blended with an Fe
{\small I} line; we also excluded this line. The line strengths and shapes for
Eu {\small II} $\lambda$3819 and $\lambda$4129 are strongly dependent on the
isotope ratio (Eu$^{151}$:Eu$^{153}$), but the appropriate value is unknown.
The solar-system isotope ratio is Eu$^{151}$:Eu$^{153}$=0.49:0.51, but most
solar-system Eu originates in the $r$-process, whereas LP~625-44 and LP~706-7
are $s$-process-dominated. \citet{kappeler89} derived the isotope ratio of the
$s$-process main component as Eu$^{151}$:Eu$^{153}$=0.02:0.98, very different
from the solar-system ratio, but the value predicted by a recent calculation
of the $s$-process in AGB stars \citep{arlandini99} is
Eu$^{151}$:Eu$^{153}$=0.54:0.46, similar to the solar-system one. Thus it is
difficult at present to know the appropriate isotope ratio for the abundance
analysis of Eu {\small II} $\lambda$3819 and $\lambda$4129 lines. For this
reason we used only two lines (Eu {\small II} $\lambda$3907 and $\lambda$4522)
for which the effects of hyperfine splitting and isotope shift are much smaller
(see Table~\ref{tab:ew}).

\paragraph{Gadolinium:} For the $\lambda$4085.5 line, the transition
probability determined by \citet{bergstrom88} was adopted. For other lines we
adopted $gf$-values in \citet{cb62} with no correction, because
\citet{bergstrom88} reported that there is no systematic discrepancy between
their results and those of \citet{cb62}.  Insufficient data were found to
permit calculations of isotopic shifts, or of hyperfine splitting for the 30\% of odd-$N$ isotopes.

\paragraph{Dysprosium:} Transition probabilities of Dy {\small II} lines were
determined by \citet{kusz92}. Since we found no useful data on hyperfine
splitting for Dy
{\small II} lines, we neglected it, but the effect should be small, because the
lines used for abundance analysis are quite weak. Spectrum synthesis was
applied because the lines have some blending with other elements.

\paragraph{Erbium, Thulium, Hafnium:} For these three elements, the line
data compiled by \citet{sneden96} were used. The abundances were
determined by spectrum synthesis with a careful check of line
blending.

A standard abundance analysis, based on the measured equivalent widths, was
applied to unblended lines, for which the equivalent widths are listed without a
dagger in Table ~\ref{tab:ew}. The standard analysis was also applied to lines
of La, Ce, Pr, Nd, and Sm, which are affected by hyperfine splitting and
isotope shifts.
In those cases, however, equivalent widths were computed by integrating the
synthetic spectrum of multi-component lines, and then comparing with the
observed ones.  As for the Pb {\small I} line, we have confirmed that there are
no distinct features of CH and CN at these wavelengths in the spectrum of
CS~22957-027, which has strong molecular features but almost no absorption by
neutron-capture elements.  Spectrum synthesis was applied in the case of lines
that are blended with lines of other elements or molecules.  In
Figure~\ref{fig:sp_heavy}, examples of the comparison between observed and
synthetic spectra for Dy {\small II}, Er {\small II} and Hf {\small II} are
shown.

The abundances derived for LP~625-44 are given in Table
\ref{tab:res}. The errors were estimated following the treatment of
\citet{ryan96}. The errors from the uncertainties of the atmospheric
parameters were evaluated by adding in quadrature the individual errors
corresponding to $\Delta T_{\rm eff}=100$K, $\Delta \log g=0.3$, and $\Delta
v=0.5$km s$^{-1}$. The internal errors were estimated by assuming the random
error in the measurement of equivalent widths to be 4 m\AA \ (and 6 m\AA \ for
Ba {\small II} in the red region) as measured in Paper I, and taking the random
error associated with uncertain $gf$ values to be 0.1 dex.  For convenience the
abundances of LP~706-7 determined by \citet{norris97a} are also listed in Table
\ref{tab:res} along with the Pb abundance derived in the previous section.

\subsection{Carbon and Nitrogen}\label{sec:cn}

Carbon, nitrogen, and oxygen abundances, and their isotope ratios, are
quite important for understanding the processes that take place in the interior
of AGB stars. The excess carbon in the outer atmospheres of AGB stars is
recognized as a result of the triple-$\alpha$ process in the thermal pulse and
mixing during the third dredge-up. Nitrogen is synthesized through the
CN(O)-cycle, while the carbon abundance decreases by this process. The carbon
isotope ratio (C$^{12}$/C$^{13}$ ratio) also usually decreases towards the
equilibrium value $\sim 3$. In the present work, the carbon and nitrogen
abundances were re-determined for LP~625-44 using the new, higher
quality-spectrum. Moreover, the carbon isotope ratio ($^{12}$C/$^{13}$C) was
also determined from the $^{13}$CH lines in the spectrum of LP~625-44.

The carbon and nitrogen abundances were determined by the molecular features of
CH at 4323\AA \ and CN at 3883\AA \, as in \citet{norris97a}. The oxygen
abundance assumed in the analysis is [O/Fe]=1.0. The derived carbon and
nitrogen abundances are not sensitive to this assumption in very carbon-rich
subgiants; we tested the range $0.0<$[O/Fe]$<1.5$, but found the effect on
abundance determination for carbon and nitrogen is negligible. One reason for
this result is that the temperature of LP~625-44 is high, and the fraction of
carbon bound in the CO molecule is quite small. Another is that the oxygen
abundance assumed here ($\log \epsilon({\rm O}) <7.67$) is much smaller than
the carbon abundance ($\log \epsilon({\rm C}) \sim8.0$), and the influence of
the assumed oxygen abundance on determination of carbon abundance is smaller
than that of oxygen-rich case. The result is given in Table~\ref{tab:res}. The
carbon abundance derived here agrees with that by \citet{norris97a} within the
uncertainty. However, the nitrogen abundance reported here is lower by 0.65~dex
than that of \citet{norris97a}. One reason is the higher carbon abundance (by
0.15~dex) inferred here, which increases the formation of CN molecules. This
explains 0.15~dex of the discrepancy. Another reason is that the dissociation
energy of the CN molecule adopted in this analysis (7.85~eV, \citet{aoki97})
is higher by 0.09~eV than that in \citet{norris97a}, and likewise increases CN
formation. As a result, the derived nitrogen abundance is lower by 0.10~dex
than that calculated assuming $E^{0}_{0}$(CN)=7.66~eV.  The oscillator
strengths of CN lines are also uncertain, as discussed in \citet{norris97a}. In
the present analysis, the oscillator strengths determined by
\citet{baushlicher88} were adopted. Because of these uncertainties in the
analysis, further investigation of other features, such as the NH
$\lambda$3360~\AA\ lines, is indispensable for a detailed discussion of 
nitrogen abundance.

The carbon isotope ratio of LP~625-44 was determined using the $^{13}$CH
features around 4200{\AA}. The synthetic spectra for $^{12}$C/$^{13}$C = 10, 20,
and 40, fitted to the observed spectrum around 4210-4225~{\AA}, are shown in
Figure~\ref{fig:sp_ch}. In this wavelength region 6 ``almost clean'' $^{13}$CH
lines are identified. The line positions were calculated using the molecular
constants derived by \citet{zachwieja95} and \citet{zachwieja97} for $^{12}$CH
and $^{13}$CH ($A^{2}\Delta$-$X^{2}\Pi$ system), respectively. In the figure we
also show the corresponding $^{12}$CH lines, which lie about 0.35~{\AA}
blueward of the $^{13}$CH lines. From the analysis, a ratio $^{12}$C/$^{13}$C
$\sim$ 20 was derived for LP~625-44. The spectrum of LP~706-7 was reviewed, but
CH features are much weaker than those in the LP~625-44 spectrum, and no useful
$^{13}$CH line was found. A higher quality spectrum is required for the
determination of the carbon isotope ratio in LP~706-7.

\section{Discussion}\label{sec:disc}

\subsection{The Over-abundance of Neutron-Capture Elements}\label{sec:disc1}

In Figure~\ref{fig:ab} the relative abundances of neutron-capture elements
([X/H]) are shown as a function of atomic number for the two stars.  The
horizontal lines indicate the values of [Fe/H].

Abundance studies of metal-poor stars with [Fe/H] $<-2.5$ have
revealed that the contribution of the $s$-process to the abundance of
neutron-capture elements is small at this level of enrichment.  This
is probably because the {\it r}-process elements originate from
nucleosynthesis in massive stars, which evolve quickly ($\sim 10^7$
yr) and eject heavy elements into the interstellar medium.  There are
almost no {\it s}-process elements formed in the early Galaxy until
the average metallicity [Fe/H] $\gtrsim -2$
\citep{mathews92}. However, we found strong excesses of
neutron-capture elements in the two metal-deficient satrs LP625-44
and LP706-7 with [Fe/H]$= -2.7$ and $-2.74$, respectively, which are
interpreted as the result of {\it s}-process nucleosynthesis from a
single site.  Namely, the abundant material polluted by {\it
s}-process nucleosynthesis dominates over the original surface
abundances of neutron-capture elements. For instance, the Ba abundance
in these two stars is a factor of several hundred times higher than
the general trend of model predictions at [Fe/H]$=-2.7$. Even the
abundance of Eu, which is usually interpreted as a signature of the
$r$-process, but should also be produced by the {\it s}-process as
well, is enhanced by more than a factor of 10 in these two stars.
Therefore, the neutron-capture elements in these two stars should
present almost pure products of $s$-process nucleosynthesis at low
metallicity. The exceptions to this are the abundances of Sr and Y in
LP~706-7, which show no distinct excess.  Therefore, the contribution
of the {\it s}-process to these two elements may not be significant
for this star.

In the following discussion, we analyse the abundance ratios of
neutron-capture elements in these two stars. We treat them as having been
produced in a single {\it s}-process site, and seek to understand investigate
the characteristics of a process that best explains the abundances of LP~625-44
and LP~706-7.

Our discussion below is not based on any specific stellar model, but
interest is focussed mostly on the nucleosynthesis process in AGB
stars (Section \ref{sec:intro} and \ref{sec:atmos}).  For
completeness, we note here some possible alternative {\it s}-process
to AGB stars. The {\it s}-process nucleosynthesis in
helium-core-burning massive stars has been studied by \citet{the00}
and \citet{rayet00} using updated neutron-capture cross sections. The
nucleosynthesis products in these models, however, have their
abundance peak centered on the lighter elements ($Z\lesssim 90$) for
any parameter choice; furthermore, massive stars are not appropriate
as the {\it s}-process site to explain our two
stars. \citet{schlattl01} pointed out a possibility of {\it s}-process
nucleosynthesis during the phase of helium core flash in low-mass,
extremely metal-poor stars based on their model calculations for the
metal-free case. This kind of study can be an useful approach to
understand the abundances of some carbon-rich and {\it s}-process-rich
stars, and should be given further attention in the future.  An
interesting recent observational result has been reported by
\citet{preston01}.  These authors conducted a long-term radial
velocity monitoring program for carbon-enhanced metal-poor stars, and
found that {\it none} of the three carbon-rich subgiant CH stars they
studied exhibited velocity variations over an 8 year period. They
conclude that these stars are likely to have undergone an enhanced
mixing event at the end of their giant-branch evolution that puts
these stars at the base of the subgiant branch again, due to increased
hydrogen mixing into their cores.  One of our stars, LP~625-44,
clearly shows a variation of radial velocity
\citep{aoki00}, and may not be similar to the stars in \citet{preston01}.
However, there is no evidence of binarity for the other, LP~706-7
\citep{norris97a}. This indicates that the suggestion by \citet{preston01}
might indeed apply to this star, and further investigation of abundances and
binarity for these (and other) subgiants, as well as the theoretical studies,
is desirable.

\subsection{Physical Conditions for {\it s}-process Nucleosynthesis}

There have been many theoretical studies of {\it s}-process nucleosynthesis in
low- to intermediate-mass AGB stars.  The best candidate reactions for the
neutron source are either the $^{13}$C($\alpha$, n)$^{16}$O
\citep{cameron55, reeves66, mathews85} or $^{22}$Ne($\alpha$,
n)$^{25}$Mg \citep{cameron60}.  In order for the former reaction to occur in
AGB stars, sufficient protons must be injected into the $^{12}$C-rich layer,
which lies below the hydrogen-rich envelope in AGB stars.  $^{13}$C is then
produced in the CN-cycle %
\begin{equation}
^{12}{\rm C}(p,\gamma)^{13}{\rm N}(e^+ \nu)^{13}{\rm C},
\label{eqn:cn}
\end{equation}
and $^{22}$Ne is produced (after the accumulation of $^{14}$N
from the CNO cycle) via the reaction sequence
\begin{equation}
^{14}{\rm N}(\alpha,\gamma)^{18}{\rm F}(e^+ \nu)^{18}{\rm
O}(\alpha,\gamma)^{22}{\rm Ne}.
\label{eqn:ne}
\end{equation}
Unfortunately, however, the precise mechanism for chemical mixing of protons
from the hydrogen-rich envelope into the $^{12}$C-rich layer is still unknown,
even for stars with solar metallicity, despite several theoretical efforts
\citep{herwig97, langer99}.  This makes it even harder to understand the
peculiar abundance pattern of the $s$-process elements found in carbon-rich,
metal-deficient stars such as LP~625-44 and LP~706-7. \citet{gallino98} and
\citet{busso99} have recently proposed an $s$-process model for metal-deficient
stars that may proceed in the so-called $^{13}$C pocket \citep{straniero95}
during the relatively long interpulse period $\sim 10000$ yr.  Since $^{13}$C
is too scarce in ordinary hydrogen-burning ashes, they have to introduce a
freely adjustable parameter to fix the total amount of $^{13}$C. This provides
enough neutrons so that the model calculation explains the {\it s}-process
abundance distribution for stars with $-2\le$[Fe/H]$\le 0$.

As an alternative, we have studied what physical conditions are necessary to
reproduce the observed {\it s}-process abundance profile of LP~625-44 and LP~
706-7 without adopting any specific stellar model.  For this purpose, we have
applied the parametric model of \citet{howard86}, with many of the
neutron-capture rates updated \citep{bao00}.  There are four parameters in this
model, only three of which are independent.  They are the neutron irradiation
time, $\Delta t$, the neutron number density, $N_{\rm n}$, the temperature,
$T_9$ (in units of $10^9$~K), at the onset of the $s$-process, and the overlap
factor, $r$, which is the fraction of material that remains to experience
subsequent neutron exposures.  These quantities can be combined to give the
neutron exposure per thermal pulse, $\tau = N_{\rm n} v_T \Delta t$, where
$v_T$ is the average thermal velocity of neutrons at $T_9$. In the case of
multiple subsequent exposures the mean neutron exposure is given by $\tau_0 =
-\tau / \ln r$.  The final abundance distributions depend only upon the neutron
exposure, as long as the neutron density is not so high that significant
branchings occur along the {\it s}-process path.  The temperature is fixed at a
reasonable value for the $^{13}$C($\alpha$,n)$^{16}$O reaction, $T_9 = 0.1$,
for these studies.  We carried out $s$-process nucleosynthesis calculations to
individually fit the abundance profile observed in LP~625-44 and LP~706-7, in
order to look for the minimum $\chi^2$ in the three-parameter space formed by
$\Delta t$, $N_{\rm n}$, and $r$.  The adopted initial abundances of seed
nuclei lighter than the iron peak elements were taken to be the solar-system
abundances, scaled to [Fe/H] = $-2.7$.  For the other heavier nuclei we use
solar-system $r$-process abundances \citep{arlandini99}, normalized to that
expected for a star with [Fe/H]$ = -2.7$.  This is a natural assumption,
because the neutron-capture-element component of the interstellar gas that
formed very metal-deficient stars is expected to consist of mostly pure {\it
r}-process elements, as proposed by \citet{truran81} and seen in various
halo-star observations \citep{spite78,gilroy88}.

Figures~\ref{sproc625} and \ref{sproc706} show our calculated
best-fit model for our two metal-deficient stars.  The parameters
deduced for LP~625-44 are $N_{\rm n}=10^7$ ${\rm cm}^{-3}$, $r=0.1$, and
$\Delta t \approx 1.7\times 10^4$ yr, which corresponds to a neutron
exposure per pulse of $\tau = 0.71 \pm0.08(1\sigma)~{\rm mb}^{-1}$ and
a mean neutron exposure $\tau_0 = (0.58
\pm0.06)\left(\frac{T_9}{0.348}\right)^{1/2}~{\rm mb}^{-1}$.  We comment below
on the permissible range of $N_{\rm n}$, and uncertainty of the
adopted parameters.  The derived parameters for the other
metal-deficient star, LP~706-7, are $\Delta t \approx1.9\times 10^4$
yr with the same $N_{\rm n}$ and $r$, which corresponds to a neutron
exposure per pulse of $\tau = 0.80
\pm0.09(1\sigma)~{\rm mb}^{-1}$, and a mean neutron exposure $\tau_0 = (0.65
\pm0.07)\left(\frac{T_9}{0.348}\right)^{1/2}~{\rm mb}^{-1}$.  The relative
abundance ratios for Pb/Sr and Ba/Sr in LP~706-7 are slightly larger than
those in LP~625-44.  This small difference is accounted for by a slight
increase of neutron exposure $\tau$.  It is noteworthy, however, that these
values for the mean exposure are significantly larger than those which best fit
solar-system material, $\tau_0 =
(0.30\pm0.01)\left(\frac{T_9}{0.348}\right)^{1/2}$
\citep{kappeler89}. \citet{gallino98} found a neutron exposure
$\tau_{\rm max} \approx 0.4$ to 0.5 in their $^{13}$C pocket model for the
solar-system $s$-process abundances.  Applying this to metal-deficient stars,
\citet{busso99} predicted an extremely enhanced Pb abundance, Pb/Ba $>$ 100, much larger than the observed value, Pb/Ba $\sim$ 1, for
LP~625-44 and LP~706-7 (We define A/B = $N_A / N_B$ and $N_A$ the
number density of nucleus $A$).

We found in our nucleosynthesis calculations that, as long as the same
neutron exposure is adopted, the abundance patterns of LP~625-44 and
LP~706-7 are reproduced with equivalent reduced $\chi^2$ values, even
in extreme conditions of very high neutron density, $N_{\rm n} \gtrsim
10^{11}{\rm cm}^{-3}$.  These parameter values simulate, more or less,
the $s$-process conditions expected during the thermal pulse phase
\citep{iben77}.  Hence, although we can constrain the neutron {\it exposure}
quite well (for this class of models), we cannot distinguish easily the neutron
{\it density} for the $s$-process based solely upon these data.

\subsection{Lead Production by Large Neutron Exposure}

The abundance analyses shown in Figures~\ref{sproc625} and
\ref{sproc706} reveal three prominent peaks at Sr-Zr, Ba, and Pb in
the $s$-process element profile, corresponding to closed neutron shells with
$N$ = 50, 82, and 126. We therefore discuss the dependence of the $s$-process
yields of Pb/Ba and Ba/Sr on the neutron exposure, $\tau$.  These ratios are
useful as a means to constrain the physical conditions of the $s$-process.

An illustration of the evidence for a large exposure, multi-pulse
model is given in Figure~\ref{exposure}.  This figure shows the calculated
elemental ratios $\log$(Pb/Ba) and $\log$(Ba/Sr) as a function of the
exposure per pulse $\tau$ in a model with $r= 0.1$.  These
are compared with the observed ratios from LP~625-44.  There is only a
narrow region of overlap, $\tau = 0.71 \pm0.08(1\sigma)~{\rm
mb}^{-1}$, in which both the observed large Ba/Sr ratio and moderate
Pb/Ba ratio can be accounted for.  The lowest panel displays the
reduced $\chi^2$ value, which is calculated in our models with all
detected elemental abundances being taken into account.  There is a
deep minimum, with $\chi^2 \approx 3$, at $\tau = 0.71~{\rm mb}^{-1}$ with
1$\sigma$ error bar $\pm 0.08~{\rm mb}^{-1}$.  There is another
shallow minimum, around $\tau \approx 2.3~{\rm mb}^{-1}$, for which the
Ba/Sr ratio is close to the observed range.  However, this parameter
is excluded because $\tau$ is so large that the predicted Pb abundance,
as well as the Pb/Ba ratio, are beyond the acceptable observed range
(see the top panel in Figure~\ref{exposure}.)

The main features of this figure can be understood qualitatively.  For
moderate neutron exposure, $\tau \approx 0.1$ to $0.7~{\rm mb}^{-1}$
($\tau_0 \approx (0.08$ to
$0.6)\left(\frac{T_9}{0.348}\right)^{1/2}~{\rm mb}^{-1}$), the product
of cross-section times abundance for the $s$-process, $\sigma_A N_A$, 
can be written \citep{mathews85, kappeler89}
\begin{equation}
\sigma_A N_A = \frac{\sigma_{A-1} N_{A-1}}{1+\frac{1}{\tau_0 \sigma_A}},
\label{eqn:steady}
\end{equation}
where $\sigma_A$ is the Maxwellian-averaged neutron-capture cross-section for
nucleus A.  This product of $\sigma_A N_A$ vs. A exhibits a characteristic
step-like function in which regions of constant $\sigma_A N_A$ make sudden
drops at neutron closed-shell nuclei $^{88}$Sr, $^{138}$Ba, and $^{208}$Pb.
After the drop, the curve is again roughly constant.  Hence, we can write the
following approximate relations for $^{88}$Sr, $^{138}$Ba, and $^{208}$Pb
\begin{equation}
\sigma_{89} N_{89} = \frac{\sigma_{88} N_{88}}{1+\frac{1}{\tau_0 \sigma_{89}}}
\approx \sigma_{137} N_{137},
\label{eqn:steady89}
\end{equation}
\begin{equation}
\sigma_{138} N_{138} = \frac{\sigma_{137} N_{137}}{1+\frac{1}{\tau_0
\sigma_{138}}}.
\label{eqn:steady138}
\end{equation}
From these we deduce
\begin{equation}
\sigma_{138} N_{138} = \frac{\sigma_{88} N_{88}}
{
\left(1+\frac{1}{\tau_0 \sigma_{89}}\right)
\left(1+\frac{1}{\tau_0 \sigma_{138}}\right)
},
\label{eqn:approx138}
\end{equation}
for which
\begin{equation}
\frac{N_{Ba}}{N_{Sr}} \approx
\frac{\tau_0^2\sigma_{88}\sigma_{89}}
{
\left(1+\tau_0 \sigma_{89}\right)
\left(1+\tau_0 \sigma_{138}\right)
}.
\label{eqn:ratebasr1}
\end{equation}
Similarly, for Pb/Ba we have
\begin{equation}
\frac{N_{Pb}}{N_{Ba}} \approx
\frac{\tau_0^2\sigma_{138}\sigma_{139}}
{
\left(1+\tau_0 \sigma_{139}\right)
\left(1+\tau_0 \sigma_{208}\right)
}.
\label{eqn:ratepbba1}
\end{equation}
Equations~(\ref{eqn:ratebasr1}) and (\ref{eqn:ratepbba1}) show the
basic behavior of a roughly quadratic increase in Ba/Sr and Pb/Ba
$\propto \tau_0^2 \propto \tau^2$ displayed in
Figure~\ref{exposure}. This relation breaks down as $\tau_0 \sigma_A$
approaches unity.  For larger exposures, the conditions $\sigma_{138}
\gg \frac{1}{\tau_0}$ and $\sigma_{208} \gg \frac{1}{\tau_0}$ can be
applied to Eq.~(\ref{eqn:ratebasr1}) and Eq.~(\ref{eqn:ratepbba1}).
The abundance ratios then asymptotically reach nearly constant values,
$\frac{N_{Ba}}{N_{Sr}} \approx 
\frac{\sigma_{88}}{\sigma_{138}}
\approx 10^{+0.3}$,
and $\frac{N_{Pb}}{N_{Ba}} \approx
\frac{\sigma_{138}}{\sigma_{208}}
\approx 10^{+0.98}$.

The deviation of these ratios from Eq.~(\ref{eqn:ratebasr1}) $-$
Eq.~(\ref{eqn:ratepbba1}) is mostly due to the fact that the
single-step-function approximation breaks down.  We can explain at least what
kind of effect might cause the deviation.  At lower neutron exposure, $\tau
\lesssim 0.1~{\rm mb}^{-1}$, the increase in both of the ratios Pb/Ba and Ba/Sr
is due to the {\it s}-process from seed {\it r}-process elements.  Since the
neutron exposure is too small to affect the $s$-process from iron-peak
elements, only a {\it weak} {\it s}-process operates on seeds from the nearby
abundance peaks of the {\it r}-process elements.  In order to verify this fact
quantitatively, we have run our {\it s}-process code without the introduction
of seed {\it r}-process elements.  The result is shown by dot-dashed
curve in Figure~\ref{exposure}.  In this case even the {\it weak} $s$-process
mentioned above does not operate at low neutron exposure, $\tau \lesssim
0.1~{\rm mb}^{-1}$, so that both Pb/Ba and Ba/Sr ratios decrease monotonically
as $\tau$ decreases.  Likewise, at intermediate neutron exposures, $0.1~{\rm
mb}^{-1} \lesssim \tau \lesssim 0.7~{\rm mb}^{-1}$, the main $s$-process operates on the very
abundant iron-peak elements, as we have already discussed in this
section. As the neutron exposure increases further, $\tau \gtrsim
0.7~{\rm mb}^{-1}$, the $s$-process starts even from the Ne-Si seed
abundance peaks which we included in the present calculations.  More
Ba than Pb and more Sr than Ba are produced from these light-mass seed
nuclei, thus regulating the abundance ratios Pb/Ba and Ba/Sr from
monotonic growth at $\tau
\gtrsim 0.7 {\rm mb}^{-1}$. It is interesting to note in
Figure~\ref{exposure} that the structure seen in the ratio Ba/Sr is
shifted towards higher neutron exposure in the ratio Pb/Ba, by a factor
$\sim$ 1.5 to 2.0.  This is a natural consequence of the fact that the
{\it s}-process produces heavy nuclei from lighter seed nuclei. The
efficiency of this is proportional to the neutron exposure.

\subsection{Single Pulse or Multi Pulse ? --- A New {\it s}-Process Paradigm}

We have extensively explored the convergence of the abundance distribution of
$s$-process elements through recurrent neutron exposures.  Almost all elements,
except for Pb, were found to be made in the first neutron exposure.  Even the
lead abundance converges after about three recurrent neutron exposures.  This
is consistent with the small overlap factor, $r \approx 0.1$, deduced in our
best-fit model.  Figure~\ref{overlap} shows the calculated elemental ratios,
$\log$(Pb/Ba) and $\log$(Ba/Sr), and reduced $\chi^2$, as a function of the
overlap factor, $r$, with fixed neutron exposure $\tau = 0.71$ for LP~625-44.
The observed Pb/Ba ratio is reproduced in the few-pulse model only for a small
overlap factor, $r \lesssim 0.2$, while the Ba/Sr ratio is rather insensitive
to $r$ and allows for a wider range, $r\lesssim 0.65$.  The Pb abundance is so
sensitive to $r$ that large $r$-values ($0.2 \lesssim r$) are almost entirely
excluded, as shown in the top panel of Figure~\ref{overlap}.  This is a
characteristic feature of the {\it s}-process pattern observed in LP~625-44 and
LP~706-7.

\citet{gallino98} have found an overlap factor of $r = 0.4 \sim 0.7$ in their
standard evolution model of low-mass (3$M_{\odot}$) AGB stars at solar
metallicity.  Theoretical estimates of $r$ were reported by \citet{iben77} for
intermediate-mass ($7M_\odot$) AGB stars, taking account the core-mass
dependence. \citet{howard86} used $r = 0.285$ in their {\it s}-process
calculations with a constant neutron density, adopting a $1.16M_{\odot}$ CO
core model. They found that the {\it s}-process abundances converge after 6 to
8 pulses.  These $r$-values are based upon AGB stars with solar metallicity,
and are very different from our value $r \approx 0.1$, found for the best fit
to metal-deficient AGB stars that produced the abundance patterns of LP~625-44
and LP~706-7.

In an $s$-process scenario that invokes radiative $^{13}$C-burning (i.e., the
$^{13}$C pocket model), a small $r \sim 0.1$ may be realized if the
third dredge-up is deep enough for the $s$-processed material to be
diluted by extensive admixture of unprocessed material.  Once this
happens, no matter how many pulses may follow, the observed abundance
profile of LP~625-44 and LP~706-7 may be reproduced in the first few
interpulses, as we demonstrated in the present calculations.

Another possibility is that the $s$-process material in
metal-deficient AGB stars has experienced only a few neutron exposures
in the convective He-burning shell.  This is consistent with a newly
proposed mechanism for the $s$-process in metal-deficient AGB stars
\citep{fujimoto00}.  These authors proposed a scenario in which
the convective shell triggered by the thermal runaway develops inside
the helium layer, and penetrates into the hydrogen-rich envelope.
This carries protons to the He- and $^{12}$C-rich layers.  Once this
occurs, $^{12}$C captures proton to synthesize $^{13}$C and other
neutron-source nuclei.  The thermal runaway continues to heat material
in the thermal pulse so that neutrons produced by the
$^{22}$Ne($\alpha$,n)$^{25}$Mg reaction as well as the
$^{13}$C($\alpha$,n)$^{16}$O reaction may contribute.  Detailed
stellar evolution calculations are therefore highly desired, in order
to clarify which site is the most likely to dominate the {\it s}-process in
metal-deficient AGB stars [interpulse \citep{gallino98}, or thermal
pulse \citep{fujimoto00, iwamoto01}].

It is worth commenting on the contrasting behavior of Pb/Ba and
Ba/Sr as a function of overlap factor $r$ seen in
Figure~\ref{overlap}. Whereas [Pb/Ba] increases with higher overlap,
Ba/Sr decreases.  The former behavior may be understood as the
achievement of a higher number of captured neutrons per seed when the
overlap factor is higher, because additional neutrons will be captured
during repeat processing. This pushes the distribution of {\it s}-process
nuclei to higher atomic numbers, especially for Pb, since it is in one
sense the end-point of the $s$-process production line.  Ba/Sr must
also increase in response to this, but for Sr an additional factor
comes into play, the enhanced production of new $s$-process
nuclei just beyond the iron peak due to great abundance of iron-peak
seeds.  This source of Sr more than makes up for the processing of Sr
towards Ba, with the net effect that Ba/Sr decreases with increasing
overlap factor, in contrast to the behavior of Pb/Ba.

\subsection{Origin of Lead}

The enrichment of Pb is one of the long-standing problems in the
chemical evolution of the Galaxy. Most ($\sim 80$\%) of the Pb in the solar
system is believed to be produced by {\it s}-process nucleosynthesis. However,
the Pb abundance in the solar system cannot be explained by the main {\it
s}-process component alone.  A strong component, with a much higher neutron
exposure, has therefore been postulated
\citep[e.g., ][]{kappeler89}.

Low-mass, metal-poor AGB stars have been proposed as a site for {\it
s}-process nucleosynthesis of Pb \citep{gallino98}.  Based on the
stellar yields and on a model of the Galactic chemical evolution,
\citet{travaglio01} discussed the origin and the enrichment history
of Pb in the Galaxy.  They concluded that low metallicity, low-mass
AGB stars are the main contributors of Pb to the Galaxy.

The yields of the {\it s}-process elements, including Pb, calculated by
the Torino models are dependent on the assumed amount of $^{13}$C (the
neutron source) in the He intershell in which {\it s}-process
nucleosynthesis occurs.  However, the amount of $^{13}$C cannot, at present, be
determined theoretically, and is constrained only by observations of the
abundances of $s$-process elements.  Theoretical arguments \citep{gallino98}
and observational constraints for moderately metal-poor stars ([Fe/H] $\sim
-1$) \citep{busso99} indicate that no single value suffices, i.e. a range of
$^{13}$C source material is required. Consequently, \citet{travaglio01} adopted
a mean of the Pb production by AGB models with different amounts of $^{13}$C.
Our results for two very metal-poor stars, in which the abundance ratios of
neutron-capture elements produced by AGB stars are well-preserved, place strong
constraints on the parameters for AGB stars with [Fe/H]$\sim -2.7$. In fact,
the ratio [Pb/Ba] = $-0.19\pm 0.28$ in LP~625-44 requires a smaller amount of
$^{13}$C than that of the so-called standard model with this metallicity
\citep{ryan01}.  The ratio is slightly higher in LP~706-7, [Pb/Ba] = $+0.27\pm
0.24$.  This {\it may} indicate that a range of $^{13}$C amounts is indeed
required in the most metal-poor AGB stars, as well as for the moderately
metal-poor ones.  However, the observational errors in the present study are
sufficiently large that the difference between these two stars is only
marginally significant.  Hence, continued observational study of abundance
ratios for neutron-capture elements, in particular of Pb, in stars such as
LP~625-44 and LP~706-7 (over a range of metallicity), is indispensable.  These
studies are necessary, both to refine models of stellar structure and
evolution, and to clarify the enrichment mechanism for neutron-capture elements
in the Galaxy.

We have pointed out that there is another possibility for the synthesis of {\it
s}-process elements in the AGB stars, i.e., with nucleosynthesis taking place
during thermal pulses, as discussed in the previous section. Further studies of
this process are also important to better understand the enrichment of Pb in
the Galaxy.

\section{Summary and Concluding Remarks} 

From an analysis of high-resolution spectra, the Pb {\small I}
$\lambda$ 4057 line is detected in the {\it s}-process-rich, very metal-poor
subgiants LP~625-44 and LP~706-7. Since the overabundance of neutron-capture
elements in LP~625-44 (and possibly in LP~706-7) can be attributed to the mass
transfer from companion AGB stars, their heavy-element abundance ratios provide
a unique opportunity to investigate {\it s}-process nucleosynthesis in AGB
stars at very low metallicity. The abundance ratios Ba/Sr and Pb/Ba are
especially strong tools to constrain the parameters of classical {\it
s}-process models.  In the context of these models, we have estimated the
neutron exposure per pulse $\tau \sim 0.7$ mb$^{-1}$, a value
significantly larger than that which best fits solar-system material ($\tau
\sim 0.4$), but consistent with the values predicted by models of rather
metal-deficient AGB stars \citep{gallino98}.  However, we also found that these
abundance ratios can be explained by very high neutron density ($N_{\rm n} \sim
10^{11}$cm$^{-1}$), as well as low ones ($N_{\rm n} \sim 10^{7}$cm$^{-1}$).
Further theoretical studies of evolved stars are required to distinguish
nucleosynthesis pathways during thermal pulses from those that take place
during the interpulse phase of AGB stars, and to identify which of these two is
the more viable site for {\it s}-process nucleosynthesis at low metallicity.

To underpin these studies, accurate abundance analyses for similar {\it
s}-process-rich, metal-poor, carbon-enhanced stars are required.  In the
present study, we have extended the abundance analysis, including the effects
of hyperfine splitting and isotope shifts, to many lines of neutron-capture
elements, including Pb. These effects are important, not only in the
determination of elemental abundances, but also in the estimation of isotope
ratios of some elements, when higher resolution and higher quality spectra
become available.  Further abundance studies of neutron-capture-rich stars will
reveal the characteristics of the {\it s}-process at low metallicity, such as
its metallicity dependence, and the history of enrichment of neutron-capture
elements in the early Galaxy.

\acknowledgments

%Acknowledgements

The authors wish to thank the Australian Time Assignment Committee
 and the Director and staff of the Anglo-Australian Telescope
for the provision of research facilities.  S.G.R. records his
deep gratitude to R. Gallino, M. Busso, and C. Travaglio for numerous
discussions on this topic, and for their hospitality during a recent
visit to Torino.  S.G.R. was supported by PPARC grant
PPA/O/S/1998/00658.  W.A., H.A. and T.K. were supported in part by the
Grant-in-Aid for Science Research 10044103, 10640236 and 12047233 of
the Ministry of Education, Science, Sports, and Culture of Japan.
G.J.M. was supported by DoE Nuclear Theory Grant DE-FG02-95-ER40394 at
UND.  T.C.B. received partial support from NSF grant AST 00-98549.  T.C.B would
also like to acknowledge support from a visiting-scholar fellowship at the
National Observatory of Japan, and the hospitality of his hosts during his
visit.

\newpage

\appendix
\section{Isotope Shifts and Hyperfine Splitting of Observed Transitions}

Many of the elements beyond iron have multiple stable isotopes, giving
rise in some cases to isotope shifts that are large enough to affect
the formation of stellar spectral lines. Furthermore, if either the
atomic number $Z$ or the neutron number $N$ of an isotope is odd, then
hyperfine splitting is also possible. Isotope shifts and hyperfine
splitting affect spectral line formation in the same way.  Their
effects are negligible for genuinely ``weak'' spectral lines, where the
line strength varies linearly with the line opacity, but once a line begins to
saturate and the relationship becomes non-linear, as it is for many spectral
lines, the wavelength distribution of the opacity becomes important.  If a
spectral line consists of multiple components whose separation is comparable to
or greater than the intrinsic width of the line from natural, thermal,
collisional, and microturbulent (but not macroturbulent) broadening, then line
splitting can be important. Optical lines, even with equivalent widths small as
20~m{\AA}, can be affected if the splitting is large enough.

Because line splitting affects the strength of a stellar spectral
line, it also affects the abundances we compute. Neglect of line
splitting in the computation of a spectral line results in an underestimate
of the equivalent width, and hence an overestimate of the abundance. If
the relative intensities and separations of the components of the line
are known, it is straightforward to calculate the impact of the
splitting on the abundance. For the astronomer, the problem is a purely 
practical one: In many cases, the line separations of different isotopes
and/or the splitting coefficients associated with hyperfine structure, 
{\it for the transitions we wish to measure in stars}, are simply unknown.

The hyperfine splitting of each energy level of a transition is characterized
by the quantum number $F$, where $F$ takes the values $I+J, I+J-1, ...,
\mid I-J\mid$, $I$ being the nuclear spin quantum number and $J$ being
the electronic angular-momentum quantum number, both of which are
known. These quantum numbers determine the number of components into
which a given level is split, and also the relative intensities of the
resulting lines. These values are relatively straightforward to
obtain. The energy separations within each of the upper and lower
levels are characterized by two hyperfine splitting coefficients $A$ and $B$,
but in many cases these are no known for transitions of interest. Where they
are known, splittings can be computed as described, for example, by
\citet{mcwilliam95}.

In this appendix, we report what is known about the lines of the rare-earth
elements and several others measured in our study.  For many of the levels
involved in the transitions we observed only the $A$ constant is known. This
is not particularly troublesome, as the $B$ constant and its impact on the
level splitting is {\it usually} much smaller than $A$'s. In all cases where
the $B$ value was not available, we assumed it to be zero.  Below, we provide
tables of wavelengths and the fraction of the total $gf$ value that should be
assigned to each component. Some lines are included in this Appendix that do
not feature in Table~2. This is because not all lines were used in the final
analysis, for reasons discussed in the main text.  We nevertheless tabulate the
components we calculated.

\subsection{Lanthanum: Pure Hyperfine Splitting}

Lanthanum has only one stable isotope, $^{139}$La, and although $^{138}$La has
half-life of $1.12\times10^{11}$~years, in the solar system it accounts for
less than 0.1\%\ of the element. Consequently, the lighter isotope can be
ignored.  However, La has an odd $Z$ and hence non-zero nuclear spin
(I($^{139}$La) = ${{7}\over{2}}^+$), giving rise to hyperfine splitting.

Unfortunately, for most of the La transitions in our star, only the
lower energy levels have published $A$ values. In these cases we were
forced to assume that the $A$ value for the upper level was zero. 
This assumption, made necessary by the lack of data, means that our
calculations of hyperfine splitting for La are imperfect. However, it
is quite common (though not universal) for $A$ values of the higher energy
levels of optical lines to be lower by a factor of 5 or 10 than the lower
energy level, so the assumption may not be too troublesome.  More to
the point, although we have had to make undesirable assumptions in
order to make progress, the inclusion of lower level splittings means
that our calculations are at least expected to be closer to reality
than if we had neglected hyperfine splitting completely.  $A$ values
were taken from H\"ohle, H\"uhnermann \& Wagner (1982).  No values
were available for the 4619~\AA\ transition. New measurements of
hyperfine splitting for La, to address the lack of data, have been made
and will be published separately (Blake \& Ryan 2002).

\subsection{Cerium}

Cerium has an even $Z$ and all of the isotopes have even $N$,
so the nuclear spin and hence hyperfine splitting is
zero. Nevertheless, isotopic splitting is possible.  There are four
isotopes, though only two are significant in the solar system, with
fractions $^{140}$Ce = 88.5\% and $^{142}$Ce = 11.1\%. Brix \&
Kopfermann (1952) list the isotope splittings for nine \ion{Ce}{2}
lines, six of which occur in our spectra. They have wavelengths from
4450 -- 4628~{\AA}, and lower excitation energies, from 0.5 -- 0.9~eV.
In all cases, the $^{142}$Ce lines lie 0.011~{\AA} redward of
$^{140}$Ce. In the absence of information on the (many) remaining
lines in our list, we applied this splitting to all of our Ce lines,
and apportioned the $gf$ value 89\%\ to $^{140}$Ce and 11\%\ to
$^{142}$Ce.  This shift approximation will not be exact, but is likely
to be better than assuming no splitting at all. As such a simple
splitting scheme has been adopted, we refrain from publishing the
\ion{Ce}{2} line-list.

\subsection{Praseodymium}

Praseodymium has only a single isotope, so there is no isotope splitting, but 
the odd $Z$ leads to a non-zero nuclear spin $I = {{5}\over{2}}^+$, and non-zero
hyperfine splitting. Values of the splitting coefficient $A$ were taken from
Ginibre (1989). In some cases the splittings are comparable to the (large)
splittings of Eu~II.

\subsection{Neodymium}

Neodymium (Z=60) has seven stable isotopes, two of which have odd $N$:
$^{143}$Nd makes up 12\%\ in the solar system and
$^{145}$Nd makes up 8\%. The remaining isotopes have no hyperfine
splitting, but isotopic shifts will affect all of them.  Isotopic shifts
between $^{144}$Nd and $^{150}$Nd were taken from Blaise et
al. (1984).  The shifts of other isotopes are derived from interval
ratios, which for Nd are non-uniform. The spacing intervals from
$^{144}$Nd relative to the interval $^{144}$Nd -- $^{150}$Nd are
$\Delta$(142, 143, 144, 145, 146, 148, 150) = ($-0.24, -0.15$, 0.00,
0.06, 0.27, 0.56, 1.00) \citep{murakawa54}.  
Solar-system isotope ratios
$^{142}$Nd:$^{143}$Nd:$^{144}$Nd:$^{145}$Nd:$^{146}$Nd:$^{148}$Nd:$^{150}$Nd
= 27:12:24:8:17:6:6 were used. 

Our initial line lists were computed before Murakawa's work was identified,
using the ratios ($-0.29, -0.15$, 0.00, 0.15, 0.33, 0.66, 1.00). As the
differences are small, we did not re-analyse the stars using the revised
splitting ratios. Nevertheless, the updated table is given below.
\citet{murakawa54} also provides hyperfine splittings constants for
the $^{143}$Nd and $^{145}$Nd isotopes, but only for just over a
quarter of the lines, and even then only for their lower levels.
Within the limits of such patchy data, it appeared that the hyperfine
splitting would still be considerably less than the isotope splitting,
and without more complete information, the decision was taken to
neglect the hyperfine splitting, especially since it would affect only
20\%\ of the isotopic composition.

\subsection{Samarium}

Samarium ($Z$=62) has seven stable isotopes. Two of these, making up 29\% of the
solar-system composition of the element, have odd $N$,
but there are few data on the hyperfine
splitting, so we confined our calculations to the isotope
shifts.  We assumed the solar-system isotope ratios
$^{144}$Sm:$^{147}$Sm:$^{148}$Sm:$^{149}$Sm:$^{150}$Sm:$^{152}$Sm:$^{154}$Sm =
3:15:11:14:7:27:23.  Shifts between the $^{148}$Sm and $^{154}$Sm isotopes were
taken from Rao et al. (1990), and in four cases shifts between the $^{152}$Sm
and $^{154}$Sm isotopes were from Brix \& Kopfermann (1949, 1952). The
splitting intervals, which are very non-uniform for Sm, were from Villemoes et
al. (1995).  For seven of our lines, no isotope shift data were found.  The
line lists for the remainder are tabulated below.  For the lines at 3941~{\AA}
and 3979~{\AA}, the isotope splitting is genuinely zero.

\subsection{Europium}

Eu has only two isotopes, but as $Z$ is odd ($Z$=63), hyperfine splitting
affects both of them. The magnetic moments of the two nuclei are $\mu_{151}$ =
3.46 and $\mu_{153}$ = 1.53, and the ratio between these gives the ratio of the
hyperfine structure coefficients, $A_{151}/A_{153} = \mu_{151}/\mu_{153} =
2.26$ (e.g., Hauge 1972). That is, the hyperfine splitting of $^{151}$Eu is
approximately twice as large as the splitting of $^{153}$Eu.  $A$ coefficients
for our lines were taken from Krebs \& Winkler (1960), if necessary using the
ratio of magnetic moments to obtain coefficients for $^{153}$Eu from the
$^{151}$Eu values.  No coefficient could be found for the upper level of the
3907~\AA\ line; this was assumed to be zero (see earlier comments for La about
this assumption). The line lists are given below.

The solar-system isotope ratio is $^{151}$Eu = 49\%, $^{153}$Eu =
51\%, so we have adopted a simple 50:50 division of the $gf$
values. This resembles a pure {\it r}-process for which
$^{151}$Eu/$^{153}$Eu = 0.96, but, as discussed in the main text, the
isotope mix for a pure {\it s}-process {\it may} be almost pure
$^{153}$Eu, whose line splitting is 2.26 times less than for
$^{151}$Eu. The line list provided can be modified for unequal isotope
ratios simply by rescaling the $gf$ values. For example, a pure
$^{153}$Eu line list would be obtained by setting the $^{151}$Eu
$frac$ values in Table A5 to zero, and doubling the $^{153}$Eu values.

\subsection{Lead}

Lead ($Z$=82) has four isotopes, one of which ($^{207}$Pb) has an odd
neutron number.  Isotope-shift data were taken from Manning, Anderson,
\& Watson (1950), and $gf$ values apportioned according to the solar- system
isotope ratios
($^{204}$Pb:$^{206}$Pb:$^{207}$Pb:$^{208}$Pb=0.015:0.236:0.226:0.523).  Manning
et al. list the isotope splittings and intensities for three components of
$^{207}$Pb due to the hyperfine splitting of the 4057~\AA\ line, but provide
only the center-of-gravity (cog) for the $^{207}$Pb component of the 3739~\AA\
line; this accounts for the different numbers of components we list below.

% REFERENCES:

%Biemont, E., Karner, C., Meyer, G., Tr\"ager, F., \& zu Putlitz, G. 1982, 
%A\&A, 107, 166.

%###############
\vspace{3cm}

\begin{figure}
\caption[]{  
Comparison of the equivalent widths measured by the present analysis
with those of \citet{norris97a} }
\label{fig:compew}
\end{figure}

\begin{figure}
\caption[]{  
Spectra at Pb {\small I} 4057.8\AA \ line (dots). The solid lines
indicate the synthetic spectra for [Pb/Fe] = 2.25, 2.55 and 2.85 for
LP~625-44, and those for [Pb/Fe] = 2.0, 2.3 and 2.6 for LP~706-7.  The
effects of hyperfine splitting and isotope shifts for the Pb {\small I}
line are included in the calculation of synthetic spectra. The spectra
of HD~140283 and CS~22957-027 are shown for comparison (see text). The
synthetic spectra for [Pb/Fe]=0 are also shown for each star. }
\label{fig:sp_pb}
\end{figure}

\begin{figure}
\caption[]{  
Comparison between the observed spectrum of LP~625-44 and synthetic spectra
for [Dy/Fe]= 1.7 (top), [Er/Fe]= 2.0 (middle) and [Hf/Fe]=
2.6(bottom).  Two alternative synthetic spectra for $\Delta$[X/Fe]$=\pm0.3$ are
also shown in each panel.}
\label{fig:sp_heavy}
\end{figure}

\begin{figure}
\caption[]{  
Comparison between the observed spectrum of LP~625-44 and synthetic spectra
for C$^{12}$/C$^{13}$ = 10, 20, and 40 for 6 $^{13}$CH lines. The line
identification is given in each panel. We also show the corresponding
$^{12}$CH lines, which lie about 0.35~{\AA} blueward of the $^{13}$CH
lines.  }
\label{fig:sp_ch}
\end{figure}

\begin{figure}
\caption[]{Abundance ratios ([X/H]) as a function of atomic number for
LP~625-44 (upper panel) and for LP~706-7 (lower panel). The horizontal lines
indicate the values of [Fe/H].  }
\label{fig:ab}
\end{figure}

\begin{figure}
%\plotone{lp625.eps}
\caption{The best fit to observational results of very
 metal-deficient star LP~625-44, using the {\it s}-process nucleosynthesis
 model with neutron exposure $\tau=0.71\pm 0.08{\rm mb}^{-1}$.
 \label{sproc625}}
\end{figure}

\begin{figure}
%\plotone{lp706.eps}
\caption{The best fit to observational results for the very
 metal-deficient star LP~706-7, using the {\it s}-process nucleosynthesis
 model with neutron exposure $\tau=0.80\pm 0.09{\rm mb}^{-1}$.
 \label{sproc706}}
\end{figure}

\begin{figure}
%\plotone{fig5.eps}
\caption{Abundance ratios $\log$(Pb/Ba) (top panel),
$\log$(Ba/Sr) (middle panel), and reduced $\chi^2$ (lower panel), as a
function of the neutron exposure per pulse, $\tau$, in a model with overlap
factor $r$= 0.1.  Solid curves refer to the theoretical results, and dashed
horizontal lines refer to the observational results with errors expressed by
dotted lines.  Dot-dashed curves refer to the theoretical results calculated
without {\it r}-process elements and using elements lighter than Fe as seed
nuclei.  See text for more details.  The shaded area illustrates the allowed
region for the theoretical model.
\label{exposure}}
\end{figure}

\begin{figure}
%\plotone{fig6.eps}
\caption{The same as those in Figure~\ref{exposure},
but as a function of the overlap factor $r$. 
\label{overlap}}
\end{figure}

\newpage

\begin{table}[htbp]
\caption[]{Observations}
\label{tab:obs}
\begin{tabular}{lccccl}
\noalign{\smallskip}
\hline
\hline
\noalign{\smallskip}
Object & RA(1950) & Dec(1950) & $V$ & $B-V$ & Obs. date \\
\noalign{\smallskip}
\hline
\noalign{\smallskip}
LP~625-44 & 16:40:38 & -01:49:42 & 11.85 & 0.69 & 1996 Oct. 21 \\
LP~706-7  & 00:41:33 & -14:11:36 & 12.11 & 0.46 & 1994-1996$^{\ast}$\\
\noalign{\smallskip}
\hline
\hline
\end{tabular}
\end{table}

\noindent
$\ast$ 1994 May 29,31, 1994 Oct. 24,25, 1996 Sep. 24

\begin{table}[htbp]
\small
\caption[]{}
\label{tab:ew}
\begin{tabular}{ccccccc}
\noalign{\smallskip}
\hline
\hline
\noalign{\smallskip}
$\lambda$(\AA) & $\chi$(eV) & $\log gf $ & $W$(m\AA) & $\log \epsilon$ & $\delta(\log \epsilon)$  & reference \\
\noalign{\smallskip}
\hline
\noalign{\smallskip}
\multicolumn{6}{c}{Sr {\small II}}  \\
\noalign{\smallskip}
\hline
\noalign{\smallskip}
 4077.71 &   0.00 &    0.150 &   151.99 &   1.33 &  & 1  \\
 4161.80 &   2.94 &   -0.500 &     7.55 &   1.53 &  & 1    \\
 4215.52 &   0.00 &   -0.170 &   123.70 &   1.27 &  & 1    \\
\noalign{\smallskip}
\hline
\noalign{\smallskip}
\multicolumn{6}{c}{Y {\small II}} \\
\noalign{\smallskip}
\hline
\noalign{\smallskip}
 3818.34 &   0.13 &   -0.980 &    35.48 &   0.67 &  & 1   \\
 3950.36 &   0.10 &   -0.490 &    45.73 &   0.37 &  & 1   \\
\noalign{\smallskip}
\hline
\noalign{\smallskip}
\multicolumn{6}{c}{Zr {\small II}} \\
\noalign{\smallskip}
\hline
\noalign{\smallskip}
 4208.98 &   0.71 &   -0.460 &    39.97 &   1.27 &   & 2  \\
 4258.05 &   0.56 &   -1.130 &    16.86 &   1.20 &   & 2   \\
 4443.00 &   1.49 &   -0.330 &    13.73 &   1.25 &   & 2   \\
 4496.97 &   0.71 &   -0.810 &    27.67 &   1.30 &   & 2   \\
\noalign{\smallskip}
\hline
\noalign{\smallskip}
\multicolumn{6}{c}{Ba {\small II}} \\
\noalign{\smallskip}
\hline
\noalign{\smallskip}
 5853.69 &   0.60 &   -1.010 &   110.05 &   2.28 & $<$0.01 & 1   \\
 6141.73 &   0.70 &   -0.070 &   176.98 &   2.21 & $\le$0.01 & 1   \\
 6496.91 &   0.60 &   -0.380 &   164.53 &   2.28 & $\le$0.02 & 1   \\
\noalign{\smallskip}
\hline
\noalign{\smallskip}
\multicolumn{6}{c}{La {\small II}} \\
\noalign{\smallskip}
\hline
\noalign{\smallskip}
 3790.83$^{\ast}$ &   0.13 &    0.143 &    75.33 &   0.94 &   0.06: & 3\\
 3949.10$^{\ast}$ &   0.40 &    0.615 &   132.32 &   0.44 &   1.52: & 3\\
 3988.52$^{\ast}$ &   0.40 &    0.080 &   119.38 &   0.80 &   1.50: & 1\\
 3995.75$^{\ast}$ &   0.17 &   -0.020 &    87.57 &   0.70 &   0.81  & 1\\
 4086.71$^{\ast}$ &   0.00 &   -0.160 &    72.14 &   0.68 &   0.33: & 1\\
 4123.23$^{\ast}$ &   0.32 &    0.120 &    78.06 &   1.16 &   0.01: & 1\\
 4238.38$^{\ast}$ &   0.40 &   -0.085 &   106.21 &   0.88 &   1.52: & 3\\
 4322.51 &   0.17 &   -1.050 &    54.94 &   1.15 &   0.33  & 3\\
 4333.76$^{\ast}$ &   0.17 &   -0.160 &   125.73 &   1.21 &   1.12: & 1\\
 4429.90$^{\ast}$ &   0.23 &   -0.370 &   118.90 &   1.80 &   0.68: & 3\\
 4558.46 &   0.32 &   -1.020 &    35.52 &   1.07 &   0.03  & 3\\
 4574.88 &   0.17 &   -1.120 &    35.13 &   0.87 &   0.16  & 3\\
 4613.39 &   0.71 &   -0.467 &    38.48 &   1.01 &   0.01: & 3\\
 4662.51 &   0.00 &   -1.280 &    40.59 &   1.02 &   0.10: & 3\\
\noalign{\smallskip}
\hline
\noalign{\smallskip}
\end{tabular}
\end{table}
%\newpage
\begin{table}[htbp]
\begin{tabular}{ccccccc}
\hline
\noalign{\smallskip}
$\lambda$(\AA) & $\chi$(eV) & $\log gf $ & $W$(m\AA) & $\log \epsilon$ & $\delta(\log \epsilon)$ & reference \\
\noalign{\smallskip}
\hline
\noalign{\smallskip}
\multicolumn{6}{c}{Ce {\small II}} \\
\noalign{\smallskip}
\hline
\noalign{\smallskip}
 3855.29 &   0.52 &    0.000 &    15.95 &   0.66 & $\le0.01:$ & 4  \\
 3904.34 &   0.55 &   -0.390 &    17.60 &   1.13 & $\le0.01:$ & 4   \\
 3909.31 &   0.45 &   -0.520 &    10.71 &   0.87 & $\le0.01:$ & 4   \\
 3940.64 &   0.49 &   -0.920 &     8.86 &   1.23 & $\le0.01:$ & 4   \\
 3980.88 &   0.71 &    0.030 &    15.89 &   0.80 & $\le0.01:$ & 4   \\
 3984.68 &   0.96 &    0.410 &    13.41 &   0.60 & $\le0.01:$ & 4   \\
 3992.91 &   0.73 &   -0.130 &    15.66 &   0.98 & $\le0.01:$ & 4   \\
 4011.56 &   0.71 &   -0.740 &    11.09 &   1.38 & $\le0.01:$ & 4   \\
 4015.88 &   1.04 &    0.000 &    15.55 &   1.17 & $\le0.01:$ & 4   \\
 4062.22 &   1.37 &    0.350 &    19.32 &   1.27 & $\le0.01:$ & 4   \\
 4065.16 &   0.90 &   -0.640 &    17.28 &   1.71 & $\le0.01:$ & 4   \\
 4070.84 &   1.53 &   -0.090 &    12.26 &   1.64 & $\le0.01:$ & 4   \\
 4076.24 &   0.81 &   -0.340 &    12.58 &   1.14 & $\le0.01:$ & 4  \\
 4117.29 &   0.74 &   -0.450 &     8.53 &   0.98 & $\le0.01:$ & 4   \\
 4119.02 &   0.55 &   -0.530 &    19.01 &   1.28 & $\le0.01:$ & 4   \\
 4148.90 &   1.09 &    0.040 &    17.94 &   1.24 & $\le0.01:$ & 4   \\
 4185.33 &   0.42 &   -0.560 &    18.32 &   1.14 & $\le0.01:$ & 4   \\
 4190.63 &   0.90 &   -0.390 &    20.68 &   1.55 & $\le0.01:$ & 4   \\
 4193.87 &   0.55 &   -0.400 &    10.77 &   0.84 & $\le0.01:$ & 4   \\
 4257.12 &   0.46 &   -1.020 &     9.45 &   1.29 & $\le0.01:$ & 4   \\
 4427.92 &   0.54 &   -0.380 &    22.01 &   1.17 & $\le0.01:$ & 4   \\
 4444.70 &   1.06 &    0.100 &    20.55 &   1.20 & $\le0.01:$ & 4   \\
 4485.52 &   0.98 &   -0.720 &    19.48 &   1.89 & $\le0.01:$ & 4   \\
 4515.86 &   1.06 &   -0.520 &     9.78 &   1.41 & $\le0.01:$ & 4   \\
 4544.96 &   0.42 &   -0.890 &    19.49 &   1.46 & $\le0.01:$ & 4   \\
 4551.30 &   0.74 &   -0.490 &    21.05 &   1.45 & $\le0.01:$ & 4   \\
\noalign{\smallskip}
\hline
\noalign{\smallskip}
\multicolumn{6}{c}{Pr {\small II}} \\
\noalign{\smallskip}
\hline
\noalign{\smallskip}
 3918.86$^{\ast}$ &   0.37 &    0.260 &    26.36 &   0.21 &   0.26  & 4 \\
 3925.47 &   0.00 &   -0.330 &    15.37 &   0.16 &   0.34  & 5 \\
 3964.26$^{\ast}$ &   0.22 &   -0.330 &    46.29 &   0.95 &   0.30  & 4 \\
 3964.81$^{\ast}$ &   0.05 &    0.090 &    52.95 &   0.38 &   0.45  & 5 \\
 3965.25$^{\ast}$ &   0.20 &   -0.130 &    54.78 &   0.78 &   0.48  & 5 \\
 3997.04$^{\ast}$ &   0.37 &   -0.100 &    22.11 &   0.62 &   0.02  & 5 \\
 4008.69$^{\ast}$ &   0.63 &    0.590 &    40.96 &   0.37 &   0.26  & 4 \\
 4056.54$^{\ast}$ &   0.63 &    0.640 &    30.87 &   0.16 &   0.28  & 4 \\
 4143.14$^{\ast}$ &   0.37 &    0.380 &    64.30 &   0.53 &   0.64  & 5\\
 4148.44 &   0.22 &   -0.720 &     5.05 &   0.25 &   0.07  & 4 \\
 4171.82 &   0.37 &   -0.340 &    21.52 &   0.68 &   0.11  & 4 \\
 4405.83$^{\ast}$ &   0.55 &   -0.350 &    59.84 &   0.94 &   0.51  & 5 \\
 4535.92 &   0.00 &   -0.980 &    15.11 &   0.73 &   0.06  & 4 \\
 4651.50 &   0.20 &   -1.030 &    14.25 &   0.95 &   0.06  & 4 \\
\noalign{\smallskip}
\hline
\noalign{\smallskip}
\end{tabular}
\end{table}
%\newpage
\begin{table}[htbp]
\begin{tabular}{ccccccc}
\hline
\noalign{\smallskip}
$\lambda$(\AA) & $\chi$(eV) & $\log gf $ & $W$(m\AA) & $\log \epsilon$ & $\delta(\log \epsilon)$ & reference\\
\noalign{\smallskip}
\hline
\noalign{\smallskip}
\multicolumn{6}{c}{Nd {\small II}} \\
\noalign{\smallskip}
\hline
\noalign{\smallskip}
 3780.40$^{\ast}$ &  0.47  &   -0.270 & 29.6  & 1.04  & 0.00   & 6 \\
 3941.51$^{\ast}$ &  0.06  &    0.150 & 54.0  & 0.77  & 0.03   & 6  \\
 3979.49$^{\ast}$ &  0.20  &   -0.110 & 40.9  & 0.83  & 0.01   & 6  \\
 4061.09$^{\ast}$ &  0.47  &    0.300 & 63.1  & 1.15  & 0.15   & 6  \\
 4133.36$^{\ast}$ &  0.32  &   -0.340 & 34.9  & 1.01  & 0.02   & 6  \\
 4446.39$^{\ast}$ &  0.20  &   -0.630 & 39.0  & 1.23  & 0.02   & 7  \\
 4451.57$^{\ast}$ &  0.38  &    0.020 & 54.6  & 1.04  & 0.12   & 6  \\
 4462.99$^{\ast}$ &  0.56  &   -0.070 & 58.3  & 1.39  & 0.15   & 6  \\
\noalign{\smallskip}
\hline
\noalign{\smallskip}
\multicolumn{6}{c}{Sm {\small II}} \\
\noalign{\smallskip}
\hline
\noalign{\smallskip}
 3922.40 &   0.38 &    0.090 &    15.49 &  -0.12 &   0.00   & 4 \\
 3941.87 &   0.00 &   -0.590 &    11.62 &   0.01 &          & 4  \\
 3979.20 &   0.54 &   -0.190 &    18.41 &   0.43 &   0.00   & 4  \\
 4220.66 &   0.54 &   -0.400 &     7.08 &   0.12 &   0.00   & 4  \\
 4244.70 &   0.28 &   -0.730 &    12.85 &   0.46 &   0.00   & 8  \\
 4318.94$^{\ast}$ &   0.28 &   -0.270 &    22.67 &   0.29 &   0.02   & 8  \\
 4424.34$^{\ast}$ &   0.49 &    0.065 &    38.20 &   0.48 &   0.08   & 4  \\
 4433.88$^{\ast}$ &   0.43 &   -0.260 &    41.08 &   0.90 &   0.00   & 4 \\
 4434.32$^{\ast}$ &   0.38 &   -0.260 &    31.00 &   0.56 &   0.05   & 4 \\
 4458.52 &   0.10 &   -0.780 &    16.19 &   0.41 &   0.02   & 4 \\
 4499.48 &   0.25 &   -1.000 &     9.01 &   0.49 &   0.00   & 8 \\
 4536.51 &   0.10 &   -1.390 &     8.20 &   0.63 &          & 4 \\
 4537.95$^{\ast}$ &   0.49 &   -0.230 &    21.02 &   0.38 &   0.04   & 8 \\
 4543.95$^{\ast}$ &   0.33 &   -0.680 &    18.48 &   0.63 &   0.00   & 4 \\
 4552.66 &   0.25 &   -1.060 &    14.38 &   0.78 &   0.00   & 4 \\
 4566.21 &   0.33 &   -0.920 &    18.09 &   0.86 &   0.00   & 4 \\
 4577.69 &   0.25 &   -0.770 &    20.61 &   0.69 &   0.00   & 8 \\
 4584.83 &   0.43 &   -0.750 &    22.37 &   0.92 &   0.00   & 4 \\
 4593.54 &   0.38 &   -0.980 &    16.83 &   0.92 &          & 4\\
 4595.29 &   0.49 &   -0.710 &    11.17 &   0.55 &          & 4\\
 4642.24$^{\ast}$ &   0.38 &   -0.520 &    32.31 &   0.84 &   0.03   & 8 \\
 4674.60 &   0.18 &   -0.560 &    15.04 &   0.22 &   0.01   & 4 \\
 4687.18 &   0.04 &   -1.170 &    19.47 &   0.78 &   0.05   & 4 \\
\noalign{\smallskip}
\hline
\noalign{\smallskip}
\end{tabular}
\end{table}
%\newpage
\begin{table}[htbp]
\begin{tabular}{ccccccc}
\hline
\noalign{\smallskip}
$\lambda$(\AA) & $\chi$(eV) & $\log gf $ & $W$(m\AA) & $\log \epsilon$ & $\delta(\log \epsilon)$ & reference\\
\noalign{\smallskip}
\hline
\noalign{\smallskip}
\multicolumn{6}{c}{Eu {\small II}} \\
\noalign{\smallskip}
\hline
\noalign{\smallskip}
% 3819.69 &   0.00 &    0.491 &     syn &    &     & 1 \\
 3907.10 &   0.21 &    0.196 &     52.0$^{\dagger}$ & -0.15 & $\le 0.15$    & 1  \\
% 4129.70 &   0.00 &    0.200 &     syn &    &     & 1  \\
 4522.57 &   0.21 &   -0.678 &     12.5$^{\dagger}$ & -0.30 & $\le 0.1$    & 1  \\
\noalign{\smallskip}
\hline
\noalign{\smallskip}
\multicolumn{6}{c}{Gd {\small II}} \\
\noalign{\smallskip}
\hline
\noalign{\smallskip}
 3844.58 &   0.14 &   -0.510 &    19.33 &   0.53 &  & 4     \\
 3957.67 &   0.60 &   -0.220 &    25.25 &   0.87 &  & 4    \\
 4070.29 &   0.56 &   -0.510 &     9.05 &   0.53 &  & 4    \\
 4073.20 &   0.43 &   -0.700 &    23.41 &   1.11 &  & 4    \\
 4085.56 &   0.73 &    0.070 &    17.13 &   0.47 &  & 9    \\
 4215.02 &   0.43 &   -0.580 &    14.35 &   0.69 &  & 4    \\
\noalign{\smallskip}
\hline
\noalign{\smallskip}
\multicolumn{6}{c}{Dy {\small II}} \\
\noalign{\smallskip}
\hline
\noalign{\smallskip}
 3757.37 &   0.10 &   -0.140 &     27.6$^{\dagger}$ &   0.1 & & 10  \\
 3944.68 &   0.00 &    0.075 &     45.8$^{\dagger}$ &   0.2 & & 10    \\
 3996.69 &   0.59 &   -0.180 &     16.3$^{\dagger}$ &   0.3 & & 10    \\
 4077.96 &   0.10 &   -0.025 &     22.4$^{\dagger}$ &  -0.2 & & 10    \\
\noalign{\smallskip} 
\hline
\noalign{\smallskip}
\multicolumn{6}{c}{Er {\small II}} \\
\noalign{\smallskip}
\hline
\noalign{\smallskip}
 3786.84 &   0.00 &   -0.640 &     29.4$^{\dagger}$ &   0.4 &  & 1   \\
 3938.63 &   0.00 &   -0.520 &     31.1$^{\dagger}$ &   0.3 &  & 1   \\
\noalign{\smallskip}
\hline
\noalign{\smallskip}
\multicolumn{6}{c}{Tm {\small II}} \\
\noalign{\smallskip}
\hline
\noalign{\smallskip}
 3700.26 &   0.03 &   -0.290 &     14.6$^{\dagger}$ &  -0.6 &  & 1   \\
\noalign{\smallskip}
\hline
\noalign{\smallskip}
\multicolumn{6}{c}{Hf {\small II}} \\
\noalign{\smallskip}
\hline
\noalign{\smallskip}
 3918.09 &   0.45 &   -1.260 &     11.0$^{\dagger}$ &   0.6 &  & 1   \\
 4093.16 &   0.45 &   -1.390 &     20.0$^{\dagger}$ &   0.9 &  & 1   \\
\noalign{\smallskip}
\hline
\noalign{\smallskip}
\multicolumn{6}{c}{Pb {\small I}} \\
\noalign{\smallskip}
\hline
\noalign{\smallskip}
 4057.815 &   1.32 &   -0.20 &     24.0$^{\dagger}$ &   1.9 & 0.1 & 11   \\
\noalign{\smallskip}
\hline
\hline
\end{tabular}
\normalsize ~ \\

\noindent
References.--- (1)\citet{sneden96}; (2)\citet{biemont81}; (3)\citet{bord96}; (4)\citet{cb62}; (5)\citet{goly91}; (6)\citet{maier77}; (7)\citet{ward85}; (8)\citet{biemont89}, (9)\citet{bergstrom88}; (10)\citet{kusz92}; (11)\citet{youssef89}

\noindent
$\ast$ Lines that were added in the present analysis to the lines studies in Paper {\small I}

\noindent
$\dagger$ Synthesized value calculated for the
abundance derived by spectrum synthesis
\end{table}

%\newpage
%\input{tab_para}

\begin{table}[htbp]
\caption[]{Stellar parameters}
\label{tab:para}
\begin{tabular}{lcccc}
\noalign{\smallskip}
\hline
\hline
\noalign{\smallskip}
Object & $T_{\rm eff}$(K) & $\log g$ & $v$(km/s) & [Fe/H] \\
\noalign{\smallskip}
\hline
\noalign{\smallskip}
LP~625-44 & 5500 & 2.8 & 1.6 & -2.71 \\
LP~706-7  & 6000 & 3.8 & 1.3 & -2.74 \\
\noalign{\smallskip}
\hline
\hline
\end{tabular}
\end{table}

\vspace{-2cm}

\begin{table}[htbp]
%\begin{center}
\caption[]{}
\label{tab:res}
\begin{tabular}{cccccccccc}
\noalign{\smallskip}
\hline
\hline
\noalign{\smallskip}
 & \multicolumn{4}{c}{LP~625-44} & & \multicolumn{4}{c}{LP~706-7} \\
\noalign{\smallskip}
\cline{2-5} \cline{7-10}
\noalign{\smallskip}
Element \hspace{2cm} & [X/Fe] & $\log\epsilon_{\rm el}$ & n & $\sigma$ & & [X/Fe] & $\log\epsilon_{\rm el}$ & n & $\sigma$ \\
\noalign{\smallskip}
\hline
Fe {\small I} ([Fe/H]) \dotfill  & $-$2.71 & 4.78 & 34 & 0.13 & & $-$2.74 & 4.75  & 74 & 0.16 \\
Fe {\small II} ([Fe/H]) \dotfill & $-$2.70 & 4.79 & 3  & 0.18 & &         &       &    &  \\
C (CH, C$_{2}$) \dotfill & +2.1    & 8.0  &    &     & & +2.15   & 7.96  &    & 0.23 \\ 
N (CN)         \dotfill  & +1.0    & 6.3  &    &     & & +1.80   & 7.03  &    & 0.35 \\ 
Sr {\small II} \dotfill  & +1.15   & 1.37 & 3  & 0.16 & & +0.15   & 0.33  & 2  & 0.18 \\
Y {\small II}  \dotfill  & +0.99   & 0.52 & 2  & 0.12 & & +0.25  & $-$0.26 & 2 & 0.19 \\
Zr {\small II} \dotfill  & +1.34   & 1.25 & 4  & 0.12 & & $<$1.16 &       & 1  & 0.21\\
Ba {\small II} \dotfill  & +2.74   & 2.26 & 3  & 0.20 & & +2.01   & 1.49  & 4  & 0.14 \\
La {\small II} \dotfill  & +2.46   & 0.98 & 14 & 0.13 & & +1.81   & 0.29  & 4  & 0.19 \\
Ce {\small II} \dotfill  & +2.27   & 1.20 & 26 & 0.12 & & +1.86   & 0.75  & 2  & 0.31 \\
Pr {\small II} \dotfill  & +2.45   & 0.58 & 14 & 0.12 & &         &       &    &      \\
Nd {\small II} \dotfill  & +2.30   & 1.08 & 8 & 0.12 & & +2.01   & 0.76  & 2  & 0.27 \\
Sm {\small II} \dotfill  & +2.21   & 0.48 & 23 & 0.12 & & $<$2.21 &       & 1  & 0.20 \\
Eu {\small II} \dotfill  & +1.97   & $-$0.2 & 2 & 0.20 && +1.40   & -0.79 & 1  & 0.20 \\
Gd {\small II} \dotfill  & +2.31   & 0.70 & 6  & 0.13 & &         &       &    &  \\
Dy {\small II} \dotfill  & +1.64   & 0.1  & 4  & 0.2  & &         &       &    &  \\
Er {\small II} \dotfill  & +2.04   & 0.3  & 2  & 0.2  & &         &       &    &  \\
Tm {\small II} \dotfill  & +1.96   & $-0.6$ & 1  & 0.2 &&         &       &    &  \\
Hf {\small II} \dotfill  & +2.76   & 0.8 & 2  & 0.2   & &         &       &    &  \\
Pb {\small I}  \dotfill  & +2.55   & 1.9  & 1  & 0.2  & & +2.28   & 1.6   & 1  & 0.2 \\
\noalign{\smallskip}
\hline
\hline
\end{tabular}
%\end{center}
\end{table}

\clearpage
\input{hfstab}

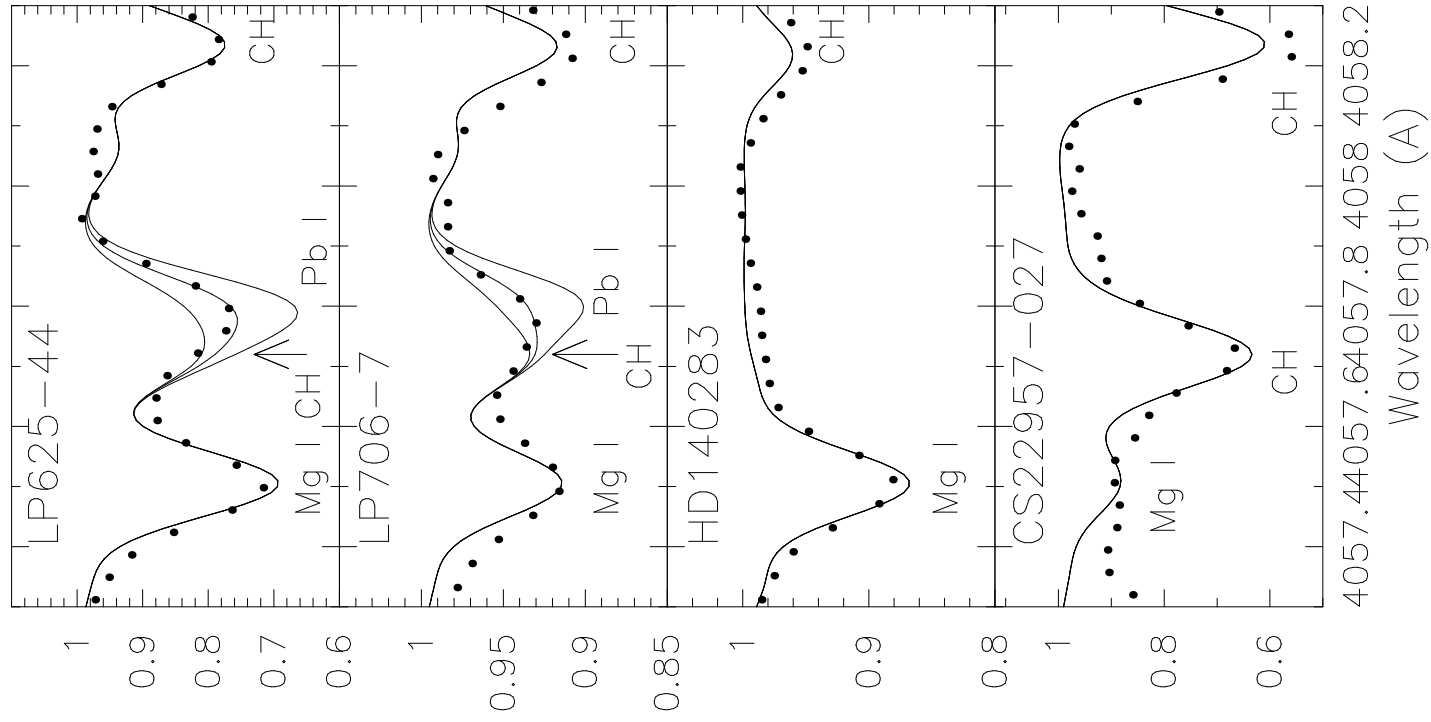
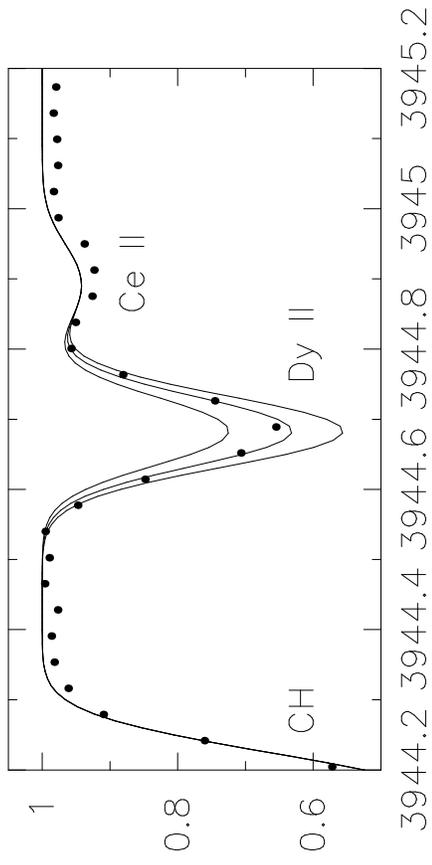
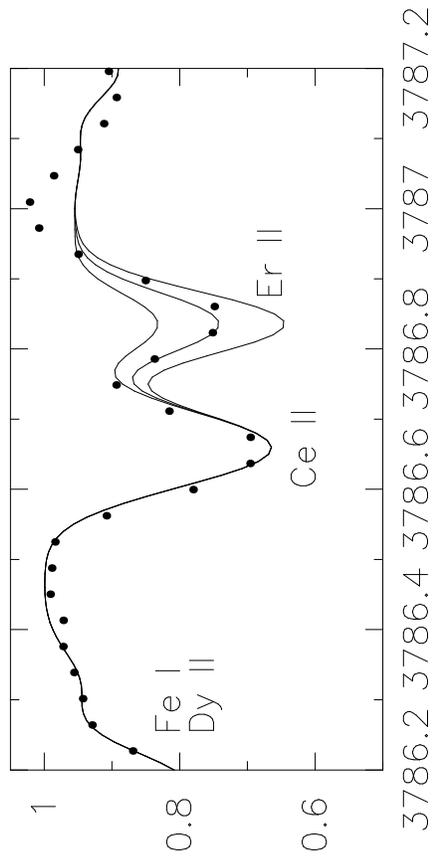
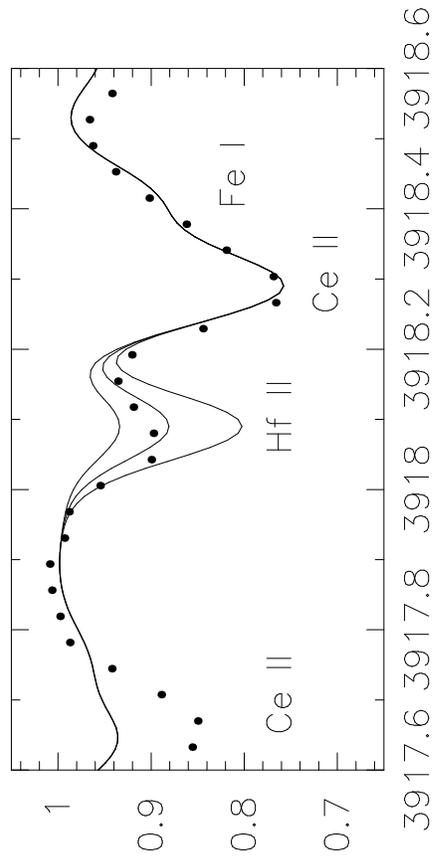
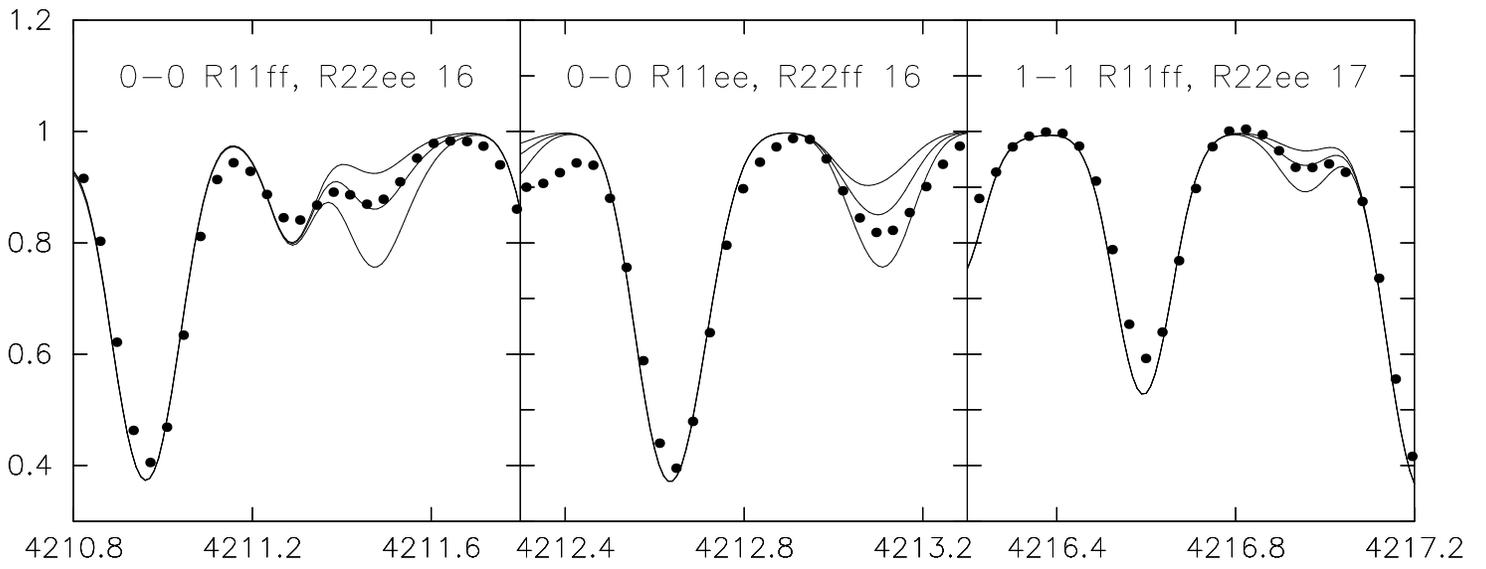
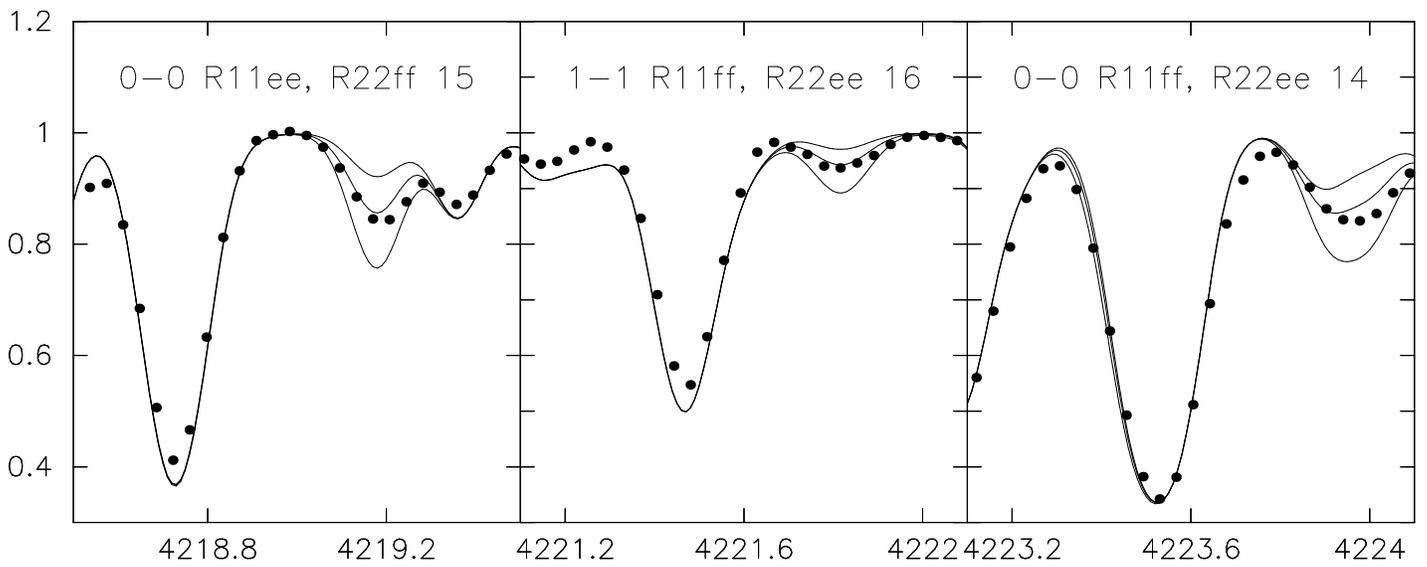
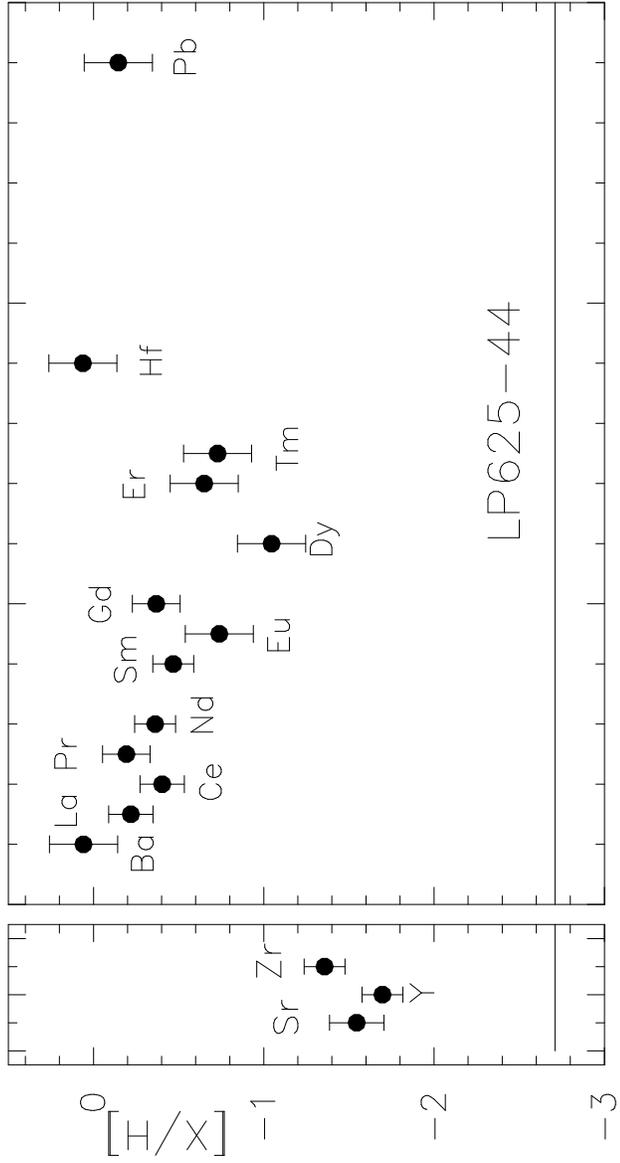
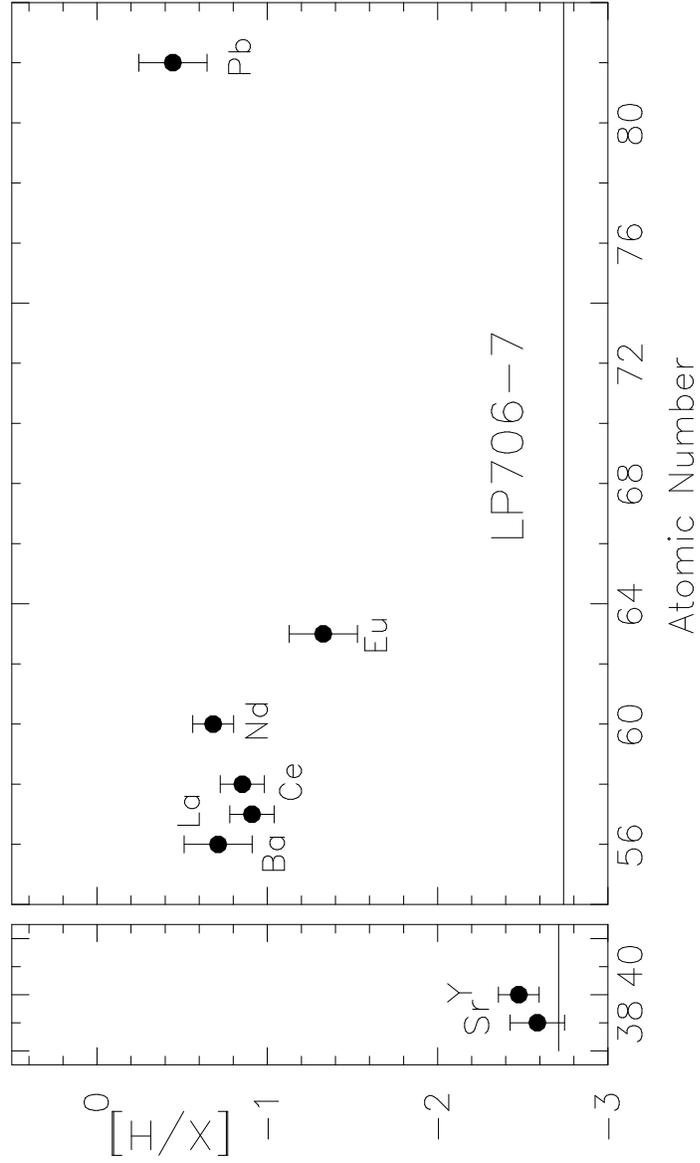
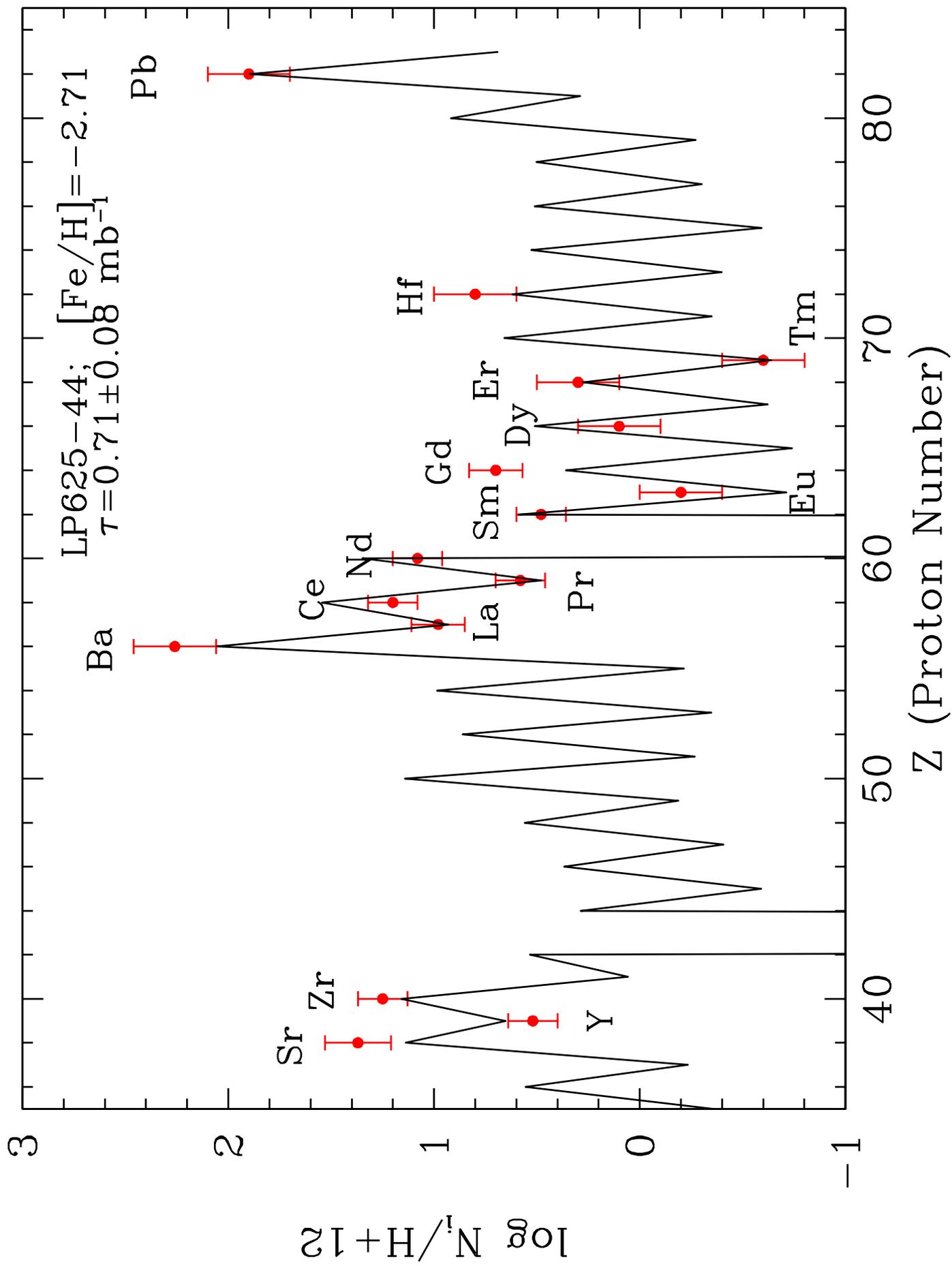
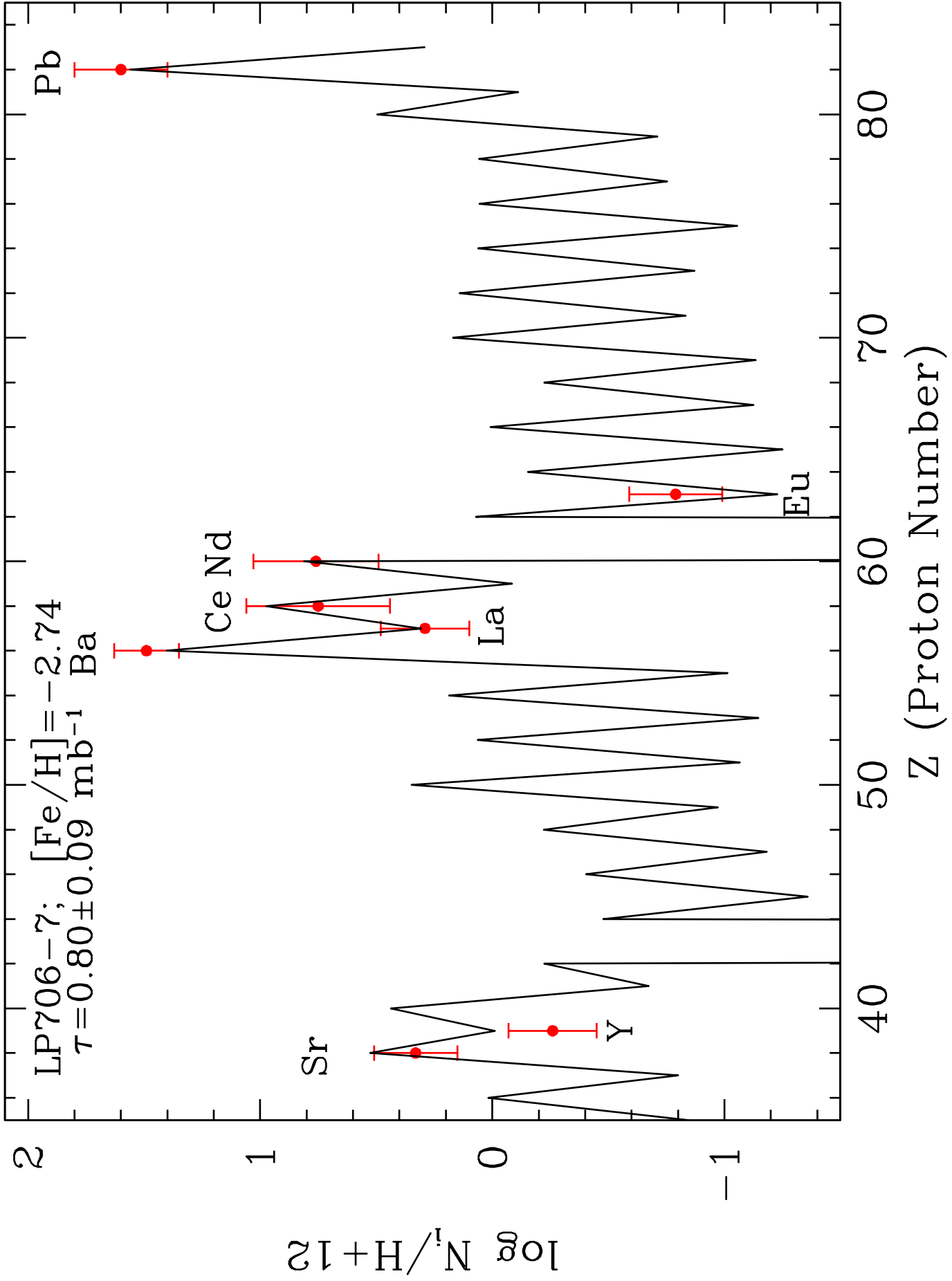
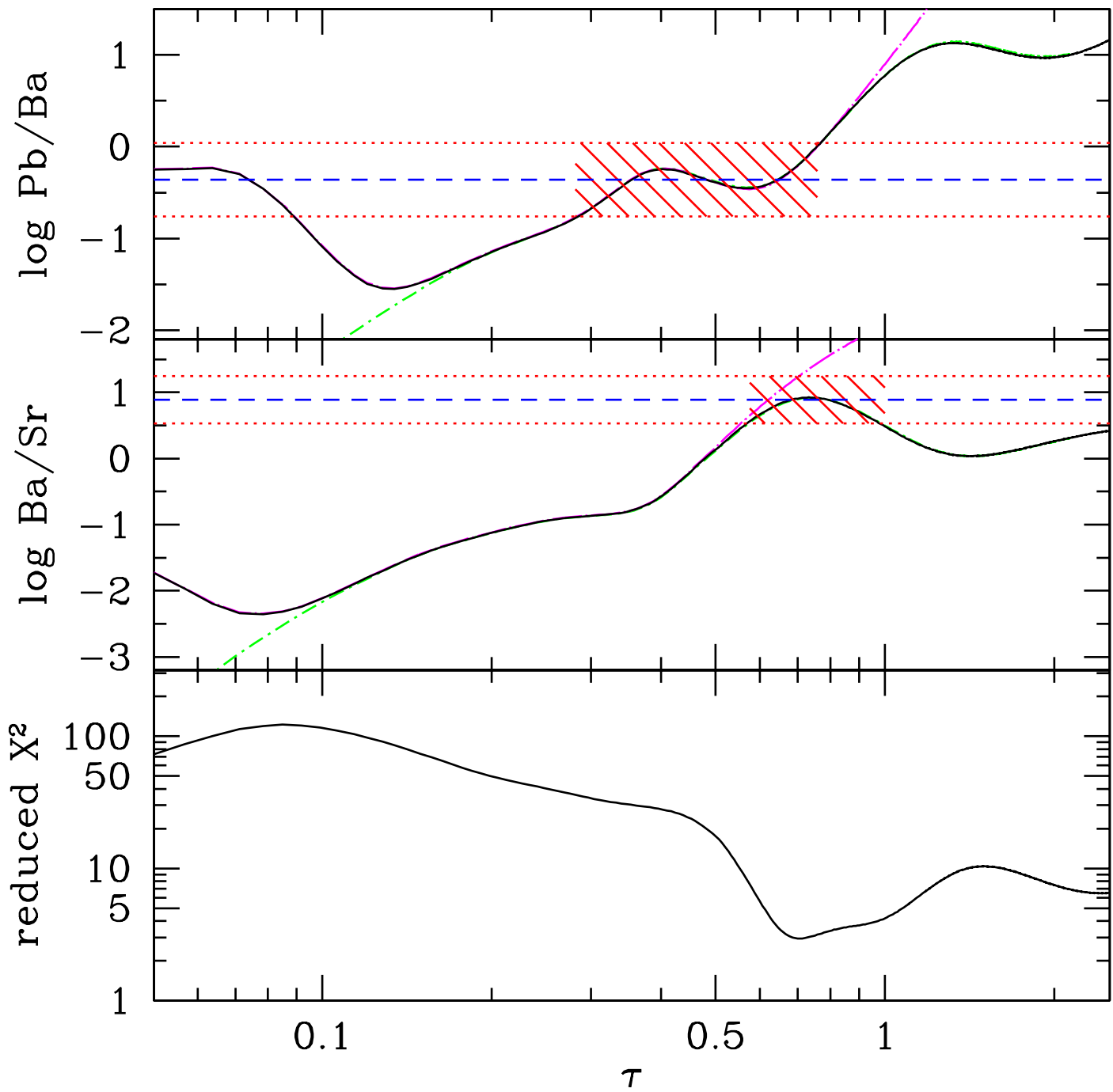
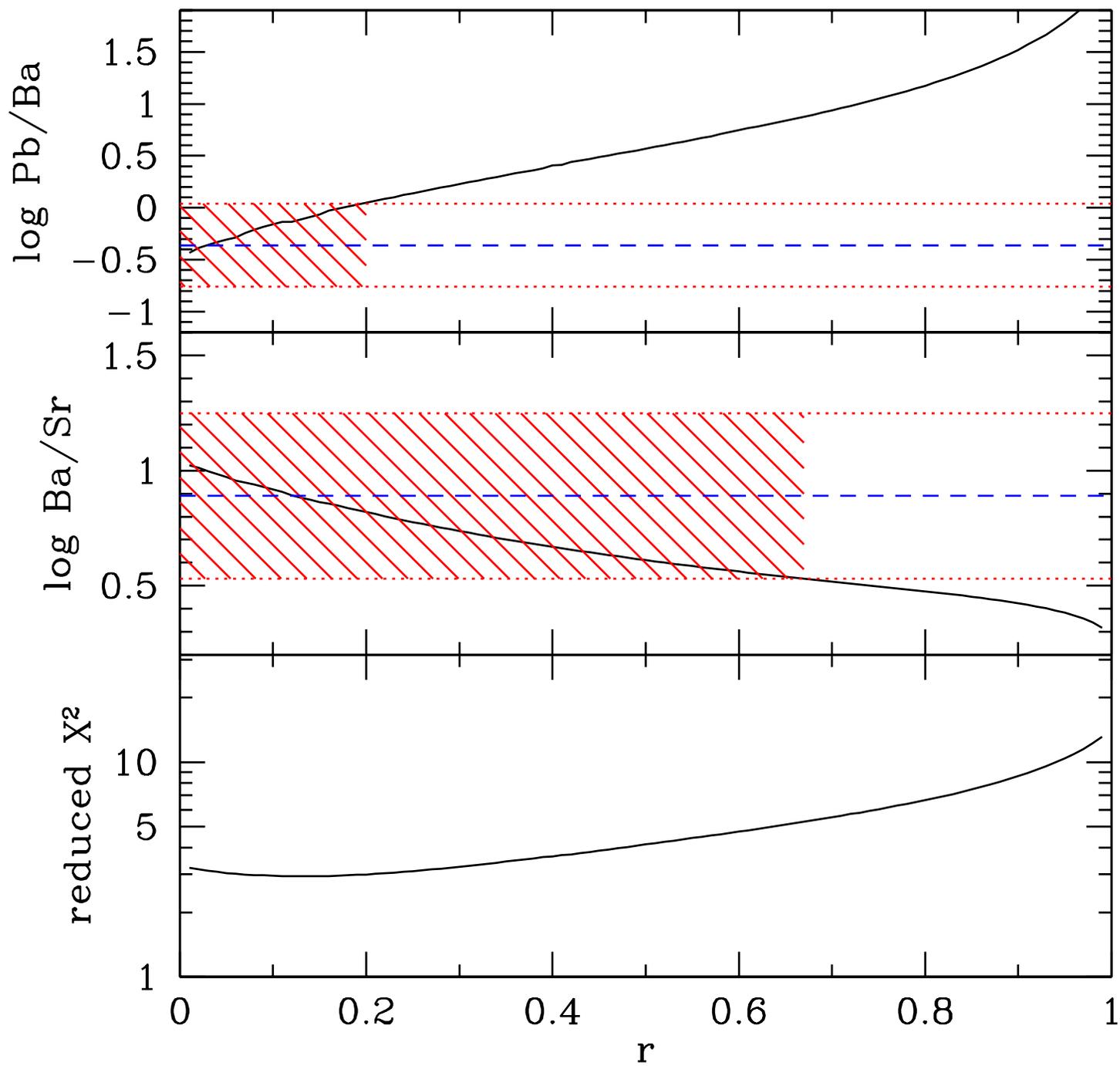
\end{document}

%% file: hfstab.tex
% TABLES:

%\documentclass[preprint]{aastex}
%\documentstyle[]{article}
%\documentstyle[apjpt4]{article}

% \voffset -50mm % this lines is needed here to position the landscape page
% \hoffset 000mm % this lines is needed here to position the landscape page

%\begin{document}
\makeatletter
\def\jnl@aj{AJ}
\ifx\revtex@jnl\jnl@aj\let\tablebreak=\nl\fi
\makeatother

\begin{center}
 
{
\small
% \bf

% \ptlandscape
% ALSO NEED TO DO   dvips $-$t landscape FILENAME   TO MAKE THE LANDSCAPE OPTION WORK PROPERLY

\begin{planotable}{lll|lll|lll|lll}
\tabletypesize{\tiny}
\tablewidth{0pt}
\tablenum{A1}
\tablecaption{Lanthanum hyperfine splitting}
\tablehead{
\colhead{$\lambda$ (\AA)\tablenotemark{a}} & \colhead{frac.} & \colhead{Label\tablenotemark{b}} &
\colhead{$\lambda$ (\AA)} & \colhead{frac.} & \colhead{Label} &
\colhead{$\lambda$ (\AA)} & \colhead{frac.} & \colhead{Label} &
\colhead{$\lambda$ (\AA)} & \colhead{frac.} & \colhead{Label} \\
}

\startdata

% 54 lines per column
3790.8237&0.0358& 0.5-1.5    &3988.5803&0.2163& 6.5-6.5    &4238.3193&0.0445& 2.5-1.5    &4558.4478&0.2503& 5.5-6.5    \\
3790.8245&0.0285& 1.5-2.5    &3988.5803&0.0338& 6.5-5.5    &4238.3413&0.0595& 3.5-4.5    &4558.4551&0.0450& 5.5-5.5    \\
3790.8245&0.0427& 1.5-1.5    &           &         &            &4238.3413&0.0270& 3.5-3.5    &4558.4551&0.1686& 4.5-5.5    \\
3790.8254&0.0205& 2.5-3.5    &3995.7131&0.0600& 1.5-2.5    &4238.3413&0.0563& 3.5-2.5    &4558.4614&0.0042& 5.5-4.5    \\
3790.8254&0.0652& 2.5-2.5    &3995.7161&0.0399& 1.5-1.5    &4238.3701&0.0528& 4.5-5.5    &4558.4614&0.0691& 4.5-4.5    \\
3790.8254&0.0215& 2.5-1.5    &3995.7212&0.0895& 2.5-3.5    &4238.3701&0.0662& 4.5-4.5    &4558.4619&0.1047& 3.5-4.5    \\
3790.8271&0.0120& 3.5-4.5    &3995.7256&0.0007& 2.5-2.5    &4238.3701&0.0595& 4.5-3.5    &4558.4663&0.0115& 4.5-3.5    \\
3790.8271&0.0755& 3.5-3.5    &3995.7288&0.0600& 2.5-1.5    &4238.4053&0.0338& 5.5-6.5    &4558.4668&0.0754& 3.5-3.5    \\
3790.8271&0.0562& 3.5-2.5    &3995.7327&0.0917& 3.5-4.5    &4238.4053&0.1278& 5.5-5.5    &4558.4673&0.0565& 2.5-3.5    \\
3790.8293&0.0045& 4.5-5.5    &3995.7388&0.0186& 3.5-3.5    &4238.4053&0.0528& 5.5-4.5    &4558.4707&0.0209& 3.5-2.5    \\
3790.8293&0.0692& 4.5-4.5    &3995.7432&0.0895& 3.5-2.5    &4238.4473&0.2163& 6.5-6.5    &4558.4712&0.0649& 2.5-2.5    \\
3790.8293&0.1047& 4.5-3.5    &3995.7478&0.0636& 4.5-5.5    &4238.4473&0.0338& 6.5-5.5    &4558.4717&0.0220& 1.5-2.5    \\
3790.8320&0.0455& 5.5-5.5    &3995.7556&0.0948& 4.5-4.5    &           &         &            &4558.4741&0.0283& 2.5-1.5    \\
3790.8320&0.1686& 5.5-4.5    &3995.7617&0.0917& 4.5-3.5    &4322.4663&0.0213& 1.5-2.5    &4558.4746&0.0429& 1.5-1.5    \\
3790.8352&0.2497& 6.5-5.5    &3995.7764&0.2363& 5.5-5.5    &4322.4692&0.0426& 1.5-1.5    &4558.4761&0.0356& 1.5-0.5    \\
         &         &            &3995.7839&0.0636& 5.5-4.5    &4322.4707&0.0359& 1.5-0.5    &           &         &            \\
3949.0256&0.0149& 0.5-1.5    &           &         &            &4322.4775&0.0561& 2.5-3.5    &4574.8223&0.0605& 1.5-2.5    \\
3949.0256&0.0209& 0.5-0.5    &4086.6899&0.0600& 1.5-2.5    &4322.4810&0.0651& 2.5-2.5    &4574.8228&0.0395& 1.5-1.5    \\
3949.0339&0.0327& 1.5-2.5    &4086.6899&0.0399& 1.5-1.5    &4322.4834&0.0281& 2.5-1.5    &4574.8379&0.0895& 2.5-3.5    \\
3949.0339&0.0318& 1.5-1.5    &4086.6956&0.0896& 2.5-3.5    &4322.4937&0.1044& 3.5-4.5    &4574.8384&0.0013& 2.5-2.5    \\
3949.0339&0.0069& 1.5-0.5    &4086.6956&0.0007& 2.5-2.5    &4322.4980&0.0752& 3.5-3.5    &4574.8389&0.0605& 2.5-1.5    \\
3949.0476&0.0562& 2.5-3.5    &4086.6956&0.0600& 2.5-1.5    &4322.5015&0.0202& 3.5-2.5    &4574.8594&0.0921& 3.5-4.5    \\
3949.0476&0.0420& 2.5-2.5    &4086.7031&0.0916& 3.5-4.5    &4322.5146&0.1684& 4.5-5.5    &4574.8604&0.0184& 3.5-3.5    \\
3949.0476&0.0089& 2.5-1.5    &4086.7031&0.0186& 3.5-3.5    &4322.5200&0.0696& 4.5-4.5    &4574.8613&0.0895& 3.5-2.5    \\
3949.0667&0.0860& 3.5-4.5    &4086.7031&0.0896& 3.5-2.5    &4322.5244&0.0123& 4.5-3.5    &4574.8877&0.0632& 4.5-5.5    \\
3949.0667&0.0484& 3.5-3.5    &4086.7131&0.0636& 4.5-5.5    &4322.5405&0.2503& 5.5-6.5    &4574.8892&0.0947& 4.5-4.5    \\
3949.0667&0.0084& 3.5-2.5    &4086.7131&0.0948& 4.5-4.5    &4322.5474&0.0460& 5.5-5.5    &4574.8901&0.0921& 4.5-3.5    \\
3949.0916&0.1231& 4.5-5.5    &4086.7131&0.0916& 4.5-3.5    &4322.5527&0.0045& 5.5-4.5    &4574.9248&0.2355& 5.5-5.5    \\
3949.0916&0.0489& 4.5-4.5    &4086.7256&0.2363& 5.5-5.5    &           &         &            &4574.9263&0.0632& 5.5-4.5    \\
3949.0916&0.0067& 4.5-3.5    &4086.7256&0.0636& 5.5-4.5    &4333.7061&0.0600& 1.5-2.5    &           &         &            \\
3949.1221&0.1682& 5.5-6.5    &           &         &            &4333.7061&0.0399& 1.5-1.5    &4613.3848&0.2997& 4.5-5.5    \\
3949.1221&0.0420& 5.5-5.5    &4123.2295&0.1686& 4.5-5.5    &4333.7207&0.0896& 2.5-3.5    &4613.3848&0.0973& 4.5-4.5    \\
3949.1221&0.0040& 5.5-4.5    &4123.2295&0.0692& 4.5-4.5    &4333.7207&0.0007& 2.5-2.5    &4613.3848&0.0196& 4.5-3.5    \\
3949.1587&0.2222& 6.5-7.5    &4123.2295&0.0120& 4.5-3.5    &4333.7207&0.0600& 2.5-1.5    &4613.3916&0.1529& 3.5-4.5    \\
3949.1587&0.0262& 6.5-6.5    &4123.2300&0.2498& 5.5-6.5    &4333.7417&0.0916& 3.5-4.5    &4613.3916&0.1272& 3.5-3.5    \\
3949.1587&0.0016& 6.5-5.5    &4123.2300&0.0454& 5.5-5.5    &4333.7417&0.0186& 3.5-3.5    &4613.3916&0.0533& 3.5-2.5    \\
         &&            &4123.2300&0.0045& 5.5-4.5    &4333.7417&0.0896& 3.5-2.5    &4613.3975&0.0533& 2.5-3.5    \\
3988.4434&0.0269& 0.5-1.5    &4123.2300&0.1047& 3.5-4.5    &4333.7686&0.0636& 4.5-5.5    &4613.3975&0.0967& 2.5-2.5    \\
3988.4434&0.0089& 0.5-0.5    &4123.2300&0.0755& 3.5-3.5    &4333.7686&0.0948& 4.5-4.5    &4613.3975&0.0999& 2.5-1.5    \\
3988.4519&0.0446& 1.5-2.5    &4123.2300&0.0205& 3.5-2.5    &4333.7686&0.0916& 4.5-3.5    &           &         &            \\
3988.4519&0.0000& 1.5-1.5    &4123.2305&0.0562& 2.5-3.5    &4333.8018&0.2363& 5.5-5.5    &4662.4839&0.0992& 1.5-2.5    \\
3988.4519&0.0269& 1.5-0.5    &4123.2305&0.0652& 2.5-2.5    &4333.8018&0.0636& 5.5-4.5    &4662.4912&0.0534& 2.5-3.5    \\
3988.4658&0.0562& 2.5-3.5    &4123.2305&0.0284& 2.5-1.5    &           &         &            &4662.4912&0.0973& 2.5-2.5    \\
3988.4658&0.0062& 2.5-2.5    &4123.2310&0.0215& 1.5-2.5    &4429.8667&0.0534& 2.5-3.5    &4662.5010&0.0191& 3.5-4.5    \\
3988.4658&0.0446& 2.5-1.5    &4123.2310&0.0427& 1.5-1.5    &4429.8667&0.0966& 2.5-2.5    &4662.5010&0.1279& 3.5-3.5    \\
3988.4856&0.0595& 3.5-4.5    &4123.2310&0.0357& 1.5-0.5    &4429.8667&0.0998& 2.5-1.5    &4662.5010&0.0534& 3.5-2.5    \\
3988.4856&0.0270& 3.5-3.5    &           &         &            &4429.8921&0.1530& 3.5-4.5    &4662.5142&0.0973& 4.5-4.5    \\
3988.4856&0.0562& 3.5-2.5    &4238.2944&0.0269& 0.5-1.5    &4429.8921&0.1273& 3.5-3.5    &4662.5142&0.1527& 4.5-3.5    \\
3988.5112&0.0528& 4.5-5.5    &4238.2944&0.0089& 0.5-0.5    &4429.8921&0.0534& 3.5-2.5    &4662.5298&0.2996& 5.5-4.5    \\
3988.5112&0.0662& 4.5-4.5    &4238.3037&0.0445& 1.5-2.5    &4429.9258&0.2997& 4.5-5.5    &           &         &            \\
3988.5112&0.0595& 4.5-3.5    &4238.3037&0.0000& 1.5-1.5    &4429.9258&0.0973& 4.5-4.5    &           &         &            \\
3988.5427&0.0338& 5.5-6.5    &4238.3037&0.0269& 1.5-0.5    &4429.9258&0.0195& 4.5-3.5    &           &         &            \\
3988.5427&0.1278& 5.5-5.5    &4238.3193&0.0563& 2.5-3.5    &           &         &            &           &         &            \\
3988.5427&0.0528& 5.5-4.5    &4238.3193&0.0063& 2.5-2.5    &           &         &            &           &         &            \\
\enddata
\tablenotetext{a}{For \ion{La}{2} $\lambda\lambda$ 3790.83, 3949.10, 3988.52, 4086.71, 4123.23, 4238.38, 4333.76, 4429.90, 4613.39,
and 4662.51, no splitting constant $A$ for the upper level was found.  $A_{\rm upper}$ was assumed zero.}
\tablenotetext{b}{F$_{\rm lower}$-F$_{\rm upper}$}
\end{planotable}
}
\end{center}

\begin{center}
 
{
\small
% \bf

% \ptlandscape
% ALSO NEED TO DO   dvips $-$t landscape FILENAME   TO MAKE THE LANDSCAPE OPTION WORK PROPERLY

\begin{planotable}{lll|lll|lll|lll}
\tabletypesize{\tiny}
\tablewidth{0pt}
\tablenum{A2}
\tablecaption{Praseodymium hyperfine splitting}
\tablehead{
\colhead{$\lambda$ (\AA)} & \colhead{frac.} & \colhead{Label\tablenotemark{a}} &
\colhead{$\lambda$ (\AA)} & \colhead{frac.} & \colhead{Label} &
\colhead{$\lambda$ (\AA)} & \colhead{frac.} & \colhead{Label} &
\colhead{$\lambda$ (\AA)} & \colhead{frac.} & \colhead{Label} \\
}
\startdata
3918.7339&0.0002& 4.5-5.5    &3964.8389&0.0007& 5.5-4.5    &4056.4663&0.2157& 9.5-10.5   &4171.7793&0.1405& 6.5-5.5    \\
3918.7517&0.0083& 4.5-4.5    &3964.8628&0.0178& 6.5-6.5    &4056.4968&0.0064& 9.5-9.5    &4171.7925&0.0132& 7.5-7.5    \\
3918.7534&0.0003& 5.5-6.5    &3964.8831&0.2308& 7.5-8.5    &4056.5054&0.1896& 8.5-9.5    &4171.8193&0.1644& 7.5-6.5    \\
3918.7664&0.0807& 4.5-3.5    &3964.8850&0.0005& 6.5-5.5    &4056.5244&0.0001& 9.5-8.5    &4171.8335&0.0084& 8.5-8.5    \\
3918.7744&0.0131& 5.5-5.5    &3964.9124&0.0113& 7.5-7.5    &4056.5330&0.0102& 8.5-8.5    &4171.8638&0.1916& 8.5-7.5    \\
3918.7769&0.0003& 6.5-7.5    &3964.9380&0.0002& 7.5-6.5    &4056.5405&0.1662& 7.5-8.5    &4171.9131&0.2222& 9.5-8.5    \\
3918.7925&0.1199& 5.5-4.5    & &         &            &4056.5576&0.0002& 8.5-7.5    & &         &            \\
3918.8013&0.0148& 6.5-6.5    &3965.1646&0.1026& 3.5-4.5    &4056.5652&0.0114& 7.5-7.5    &4405.7026&0.0057& 5.5-6.5    \\
3918.8044&0.0001& 7.5-8.5    &3965.1863&0.1199& 4.5-5.5    &4056.5720&0.1453& 6.5-7.5    &4405.7280&0.1020& 5.5-5.5    \\
3918.8223&0.1405& 6.5-5.5    &3965.2031&0.0083& 4.5-4.5    &4056.5872&0.0002& 7.5-6.5    &4405.7285&0.0090& 6.5-7.5    \\
3918.8320&0.0132& 7.5-7.5    &3965.2134&0.1405& 5.5-6.5    &4056.5938&0.0101& 6.5-6.5    &4405.7573&0.1226& 6.5-6.5    \\
3918.8564&0.1644& 7.5-6.5    &3965.2334&0.0131& 5.5-5.5    &4056.5996&0.1269& 5.5-6.5    &4405.7583&0.0101& 7.5-8.5    \\
3918.8672&0.0084& 8.5-8.5    &3965.2458&0.1644& 6.5-7.5    &4056.6128&0.0001& 6.5-5.5    &4405.7827&0.0057& 6.5-5.5    \\
3918.8948&0.1916& 8.5-7.5    &3965.2502&0.0002& 5.5-4.5    &4056.6184&0.0064& 5.5-5.5    &4405.7915&0.1378& 7.5-7.5    \\
3918.9370&0.2222& 9.5-8.5    &3965.2690&0.0148& 6.5-6.5    &4056.6233&0.1111& 4.5-5.5    &4405.7930&0.0090& 8.5-9.5    \\
         &         &            &3965.2839&0.1916& 7.5-8.5    &          &         &            &4405.8208&0.0090& 7.5-6.5    \\
3925.4172&0.2394& 6.5-6.5    &3965.2891&0.0003& 6.5-5.5    &4143.0527&0.1041& 4.5-5.5    &4405.8301&0.1573& 8.5-8.5    \\
3925.4246&0.0199& 5.5-6.5    &3965.3101&0.0132& 7.5-7.5    &4143.0757&0.1269& 5.5-6.5    &4405.8325&0.0057& 9.5-10.5   \\
3925.4417&0.0199& 6.5-5.5    &3965.3276&0.2222& 8.5-9.5    &4143.0981&0.0064& 5.5-5.5    &4405.8633&0.0101& 8.5-7.5    \\
3925.4490&0.1716& 5.5-5.5    &3965.3333&0.0003& 7.5-6.5    &4143.1035&0.1453& 6.5-7.5    &4405.8730&0.1813& 9.5-9.5    \\
3925.4551&0.0307& 4.5-5.5    &3965.3569&0.0084& 8.5-8.5    &4143.1294&0.0101& 6.5-6.5    &4405.9102&0.0090& 9.5-8.5    \\
3925.4697&0.0307& 5.5-4.5    &3965.3831&0.0001& 8.5-7.5    &4143.1362&0.1662& 7.5-8.5    &4405.9209&0.2100& 10.5-10.5  \\
3925.4758&0.1212& 4.5-4.5    & &         &            &4143.1514&0.0001& 6.5-5.5    &4405.9619&0.0057& 10.5-9.5   \\
3925.4810&0.0333& 3.5-4.5    &3996.9331&0.0072& 4.5-5.5    &4143.1650&0.0114& 7.5-7.5    &          &         &            \\
3925.4927&0.0333& 4.5-3.5    &3996.9500&0.1039& 4.5-4.5    &4143.1733&0.1896& 8.5-9.5    &4535.8579&0.2425& 6.5-7.5    \\
3925.4978&0.0857& 3.5-3.5    &3996.9553&0.0114& 5.5-6.5    &4143.1909&0.0002& 7.5-6.5    &4535.8862&0.0163& 6.5-6.5    \\
3925.5017&0.0292& 2.5-3.5    &3996.9753&0.1147& 5.5-5.5    &4143.2061&0.0102& 8.5-8.5    &4535.8960&0.1957& 5.5-6.5    \\
3925.5110&0.0292& 3.5-2.5    &3996.9822&0.0129& 6.5-7.5    &4143.2158&0.2157& 9.5-10.5   &4535.9106&0.0004& 6.5-5.5    \\
3925.5149&0.0634& 2.5-2.5    &3996.9922&0.0072& 5.5-4.5    &4143.2354&0.0002& 8.5-7.5    &4535.9204&0.0251& 5.5-5.5    \\
3925.5178&0.0184& 1.5-2.5    &3997.0051&0.1313& 6.5-6.5    &4143.2520&0.0064& 9.5-9.5    &4535.9287&0.1558& 4.5-5.5    \\
3925.5244&0.0184& 2.5-1.5    &3997.0137&0.0115& 7.5-8.5    &4143.2842&0.0001& 9.5-8.5    &4535.9409&0.0011& 5.5-4.5    \\
3925.5271&0.0558& 1.5-1.5    &3997.0251&0.0114& 6.5-5.5    & &         &            &4535.9492&0.0277& 4.5-4.5    \\
         &         &            &3997.0398&0.1534& 7.5-7.5    &4148.3887&0.2289& 7.5-7.5    &4535.9561&0.1228& 3.5-4.5    \\
3964.2068&0.2308& 7.5-8.5    &3997.0496&0.0073& 8.5-9.5    &4148.3901&0.0136& 6.5-7.5    &4535.9658&0.0015& 4.5-3.5    \\
3964.2339&0.0113& 7.5-7.5    &3997.0627&0.0129& 7.5-6.5    &4148.4175&0.0136& 7.5-6.5    &4535.9727&0.0247& 3.5-3.5    \\
3964.2354&0.1936& 6.5-7.5    &3997.0789&0.1812& 8.5-8.5    &4148.4194&0.1777& 6.5-6.5    &4535.9780&0.0954& 2.5-3.5    \\
3964.2578&0.0003& 7.5-6.5    &3997.1050&0.0115& 8.5-7.5    &4148.4209&0.0211& 5.5-6.5    &4535.9858&0.0011& 3.5-2.5    \\
3964.2593&0.0178& 6.5-6.5    &3997.1226&0.2149& 9.5-9.5    &4148.4448&0.0211& 6.5-5.5    &4535.9912&0.0156& 2.5-2.5    \\
3964.2605&0.1613& 5.5-6.5    &3997.1516&0.0073& 9.5-8.5    &4148.4463&0.1376& 5.5-5.5    &4535.9946&0.0741& 1.5-2.5    \\
3964.2800&0.0004& 6.5-5.5    & &&            &4148.4473&0.0232& 4.5-5.5    & &         &            \\
3964.2813&0.0198& 5.5-5.5    &4008.6284&0.2149& 9.5-9.5    &4148.4673&0.0232& 5.5-4.5    &4651.3594&0.0094& 3.5-4.5    \\
3964.2825&0.1336& 4.5-5.5    &4008.6367&0.0073& 8.5-9.5    &4148.4688&0.1077& 4.5-4.5    &4651.3823&0.0930& 3.5-3.5    \\
3964.2988&0.0007& 5.5-4.5    &4008.6536&0.0073& 9.5-8.5    &4148.4697&0.0205& 3.5-4.5    &4651.3848&0.0149& 4.5-5.5    \\
3964.3000&0.0175& 4.5-4.5    &4008.6619&0.1812& 8.5-8.5    &4148.4863&0.0205& 4.5-3.5    &4651.4126&0.1037& 4.5-4.5    \\
3964.3008&0.1101& 3.5-4.5    &4008.6692&0.0115& 7.5-8.5    &4148.4873&0.0877& 3.5-3.5    &4651.4165&0.0166& 5.5-6.5    \\
3964.3145&0.0004& 4.5-3.5    &4008.6843&0.0115& 8.5-7.5    &4148.4878&0.0129& 2.5-3.5    &4651.4355&0.0094& 4.5-3.5    \\
3964.3152&0.0113& 3.5-3.5    &4008.6917&0.1534& 7.5-7.5    &4148.5010&0.0129& 3.5-2.5    &4651.4497&0.1221& 5.5-5.5    \\
3964.3159&0.0909& 2.5-3.5    &4008.6982&0.0129& 6.5-7.5    &4148.5015&0.0779& 2.5-2.5    &4651.4551&0.0154& 6.5-7.5    \\
         &         &            &4008.7117&0.0129& 7.5-6.5    & &         &            &4651.4775&0.0149& 5.5-4.5    \\
3964.7188&0.0909& 2.5-3.5    &4008.7180&0.1313& 6.5-6.5    &4171.6792&0.0002& 4.5-5.5    &4651.4932&0.1477& 6.5-6.5    \\
3964.7383&0.1101& 3.5-4.5    &4008.7236&0.0114& 5.5-6.5    &4171.6987&0.0083& 4.5-4.5    &4651.5005&0.0098& 7.5-8.5    \\
3964.7539&0.0113& 3.5-3.5    &4008.7354&0.0114& 6.5-5.5    &4171.7021&0.0003& 5.5-6.5    &4651.5264&0.0166& 6.5-5.5    \\
3964.7646&0.1336& 4.5-5.5    &4008.7410&0.1147& 5.5-5.5    &4171.7144&0.1026& 4.5-3.5    &4651.5439&0.1801& 7.5-7.5    \\
3964.7834&0.0175& 4.5-4.5    &4008.7458&0.0072& 4.5-5.5    &4171.7251&0.0131& 5.5-5.5    &4651.5815&0.0154& 7.5-6.5    \\
3964.7974&0.1613& 5.5-6.5    &4008.7556&0.0072& 5.5-4.5    &4171.7300&0.0003& 6.5-7.5    &4651.6011&0.2211& 8.5-8.5    \\
3964.7991&0.0005& 4.5-3.5    &4008.7603&0.1039& 4.5-4.5    &4171.7446&0.1199& 5.5-4.5    &4651.6440&0.0098& 8.5-7.5    \\
3964.8198&0.0199& 5.5-5.5    & &&            &4171.7563&0.0148& 6.5-6.5    & &         &            \\
3964.8369&0.1936& 6.5-7.5    & &         &            &4171.7627&0.0001& 7.5-8.5    & &         &            \\
\enddata
\tablenotetext{a}{F$_{\rm lower}$-F$_{\rm upper}$}
\end{planotable}
}
\end{center}

\begin{center}
 
{
\small
% \bf

% \ptlandscape
% ALSO NEED TO DO   dvips $-$t landscape FILENAME   TO MAKE THE LANDSCAPE OPTION WORK PROPERLY

\begin{planotable}{lll|lll|lll|lll}
\tabletypesize{\tiny}
\tablewidth{0pt}
\tablenum{A3}
\tablecaption{Neodymium isotope splitting}
\tablehead{
\colhead{$\lambda$ (\AA)} & \colhead{frac.} & \colhead{Label} &
\colhead{$\lambda$ (\AA)} & \colhead{frac.} & \colhead{Label} &
\colhead{$\lambda$ (\AA)} & \colhead{frac.} & \colhead{Label} &
\colhead{$\lambda$ (\AA)} & \colhead{frac.} & \colhead{Label} \\
}
\startdata
3780.3977&0.2713&142&3911.1657&0.2713&142&4011.0858&0.2713&142&4069.2763&0.2713&  142       \\
3780.3986&0.1218&143&3911.1670&0.1218&143&4011.0874&0.1218&143&4069.2777&0.1218&  143       \\
3780.3997&0.2380&144&3911.1686&0.2380&144&4011.0894&0.2380&144&4069.2795&0.2380&  144       \\
3780.4001&0.0830&145&3911.1692&0.0830&145&4011.0903&0.0830&145&4069.2802&0.0830&  145       \\
3780.4017&0.1719&146&3911.1715&0.1719&146&4011.0932&0.1719&146&4069.2828&0.1719&  146       \\
3780.4038&0.0576&148&3911.1746&0.0576&148&4011.0972&0.0576&148&4069.2863&0.0576&  148       \\
3780.4071&0.0564&150&3911.1794&0.0564&150&4011.1033&0.0564&150&4069.2916&0.0564&  150       \\
         &         &            &          &         &            &          &         &            &          &         &            \\
3784.2483&0.2713&142&3927.0994&0.2713&142&4012.7036&0.2713&142&4085.8180&0.2713&  142       \\
3784.2489&0.1218&143&3927.0996&0.1218&143&4012.7037&0.1218&143&4085.8188&0.1218&  143       \\
3784.2498&0.2380&144&3927.0999&0.2380&144&4012.7039&0.2380&144&4085.8197&0.2380&  144       \\
3784.2501&0.0830&145&3927.1000&0.0830&145&4012.7040&0.0830&145&4085.8201&0.0830&  145       \\
3784.2513&0.1719&146&3927.1005&0.1719&146&4012.7043&0.1719&146&4085.8215&0.1719&  146       \\
3784.2530&0.0576&148&3927.1010&0.0576&148&4012.7047&0.0576&148&4085.8235&0.0576&  148       \\
3784.2555&0.0564&150&3927.1019&0.0564&150&4012.7054&0.0564&150&4085.8264&0.0564&  150       \\
         &         &            &          &         &            &          &         &            &          &         &            \\
3802.2983&0.2713&142&3941.5080&0.2713&142&4020.8664&0.2713&142&4098.1776&0.0564&  150       \\
3802.2989&0.1218&143&3941.5088&0.1218&143&4020.8678&0.1218&143&4098.1787&0.0576&  148       \\
3802.2998&0.2380&144&3941.5097&0.2380&144&4020.8695&0.2380&144&4098.1794&0.1719&  146       \\
3802.3001&0.0830&145&3941.5101&0.0830&145&4020.8702&0.0830&145&4098.1800&0.0830&  145       \\
3802.3013&0.1719&146&3941.5115&0.1719&146&4020.8727&0.1719&146&4098.1801&0.2380&  144       \\
3802.3029&0.0576&148&3941.5135&0.0576&148&4020.8761&0.0576&148&4098.1805&0.1218&  143       \\
3802.3054&0.0564&150&3941.5164&0.0564&150&4020.8813&0.0564&150&4098.1808&0.2713&  142       \\
         &         &            &          &         &            &          &         &            &          &         &            \\
3803.4730&0.2713&142&3973.2979&0.2713&142&4021.3385&0.2713&142&4109.0758&0.0564&  150       \\
3803.4734&0.1218&143&3973.2987&0.1218&143&4021.3391&0.1218&143&4109.0777&0.0576&  148       \\
3803.4739&0.2380&144&3973.2997&0.2380&144&4021.3398&0.2380&144&4109.0790&0.1719&  146       \\
3803.4741&0.0830&145&3973.3001&0.0830&145&4021.3401&0.0830&145&4109.0799&0.0830&  145       \\
3803.4747&0.1719&146&3973.3016&0.1719&146&4021.3411&0.1719&146&4109.0802&0.2380&  144       \\
3803.4756&0.0576&148&3973.3036&0.0576&148&4021.3425&0.0576&148&4109.0808&0.1218&  143       \\
3803.4771&0.0564&150&3973.3067&0.0564&150&4021.3447&0.0564&150&4109.0813&0.2713&  142       \\
         &&            &          &&            &          &&            &          &&            \\
3826.4200&0.2713&142&3973.6883&0.2713&142&4022.9975&0.0564&150&4110.4765&0.2713&  142       \\
3826.4200&0.1218&143&3973.6889&0.1218&143&4022.9996&0.0576&148&4110.4779&0.1218&  143       \\
3826.4200&0.2380&144&3973.6898&0.2380&144&4023.0009&0.1719&146&4110.4795&0.2380&  144       \\
3826.4200&0.0830&145&3973.6901&0.0830&145&4023.0019&0.0830&145&4110.4802&0.0830&  145       \\
3826.4200&0.1719&146&3973.6913&0.1719&146&4023.0022&0.2380&144&4110.4826&0.1719&  146       \\
3826.4199&0.0576&148&3973.6929&0.0576&148&4023.0029&0.1218&143&4110.4859&0.0576&  148       \\
3826.4199&0.0564&150&3973.6955&0.0564&150&4023.0034&0.2713&142&4110.4910&0.0564&  150       \\
         &&            &          &&            &          &&            &          &&            \\
3859.4190&0.2713&142&3976.8469&0.2713&142&4024.7786&0.2713&142&4116.7528&0.0563&  150       \\
3859.4194&0.1218&143&3976.8481&0.1218&143&4024.7791&0.1218&143&4116.7607&0.0576&  148       \\
3859.4199&0.2380&144&3976.8496&0.2380&144&4024.7798&0.2380&144&4116.7659&0.1720&  146       \\
3859.4201&0.0830&145&3976.8502&0.0830&145&4024.7801&0.0830&145&4116.7697&0.0831&  145       \\
3859.4208&0.1719&146&3976.8523&0.1719&146&4024.7811&0.1719&146&4116.7707&0.2381&  144       \\
3859.4217&0.0576&148&3976.8552&0.0576&148&4024.7824&0.0576&148&4116.7734&0.1219&  143       \\
3859.4231&0.0564&150&3976.8597&0.0564&150&4024.7845&0.0564&150&4116.7755&0.2711&  142       \\
         &&            &          &&            &          &&            &          &&            \\
3863.3280&0.2713&142&3979.4864&0.2713&142&4038.1192&0.0564&150&4133.3551&0.2713&  142       \\
3863.3288&0.1218&143&3979.4878&0.1218&143&4038.1196&0.0576&148&4133.3570&0.1218&  143       \\
3863.3297&0.2380&144&3979.4895&0.2380&144&4038.1198&0.1719&146&4133.3594&0.2380&  144       \\
3863.3301&0.0830&145&3979.4902&0.0830&145&4038.1200&0.0830&145&4133.3603&0.0830&  145       \\
3863.3315&0.1719&146&3979.4927&0.1719&146&4038.1200&0.2380&144&4133.3637&0.1719&  146       \\
3863.3333&0.0576&148&3979.4961&0.0576&148&4038.1202&0.1218&143&4133.3684&0.0576&  148       \\
3863.3362&0.0564&150&3979.5013&0.0564&150&4038.1202&0.2713&142&4133.3756&0.0564&  150       \\
         &&            &          &&            &          &&            &          &&            \\
3865.9782&0.2713&142&3981.2365&0.0564&150&4051.1482&0.2713&142&4160.5654&0.2713&  142       \\
3865.9789&0.1218&143&3981.2381&0.0576&148&4051.1489&0.1218&143&4160.5672&0.1218&  143       \\
3865.9798&0.2380&144&3981.2392&0.1719&146&4051.1498&0.2380&144&4160.5694&0.2380&  144       \\
3865.9801&0.0830&145&3981.2399&0.0830&145&4051.1501&0.0830&145&4160.5703&0.0830&  145       \\
3865.9813&0.1719&146&3981.2401&0.2380&144&4051.1513&0.1719&146&4160.5735&0.1719&  146       \\
3865.9830&0.0576&148&3981.2407&0.1218&143&4051.1530&0.0576&148&4160.5779&0.0576&  148       \\
3865.9856&0.0564&150&3981.2411&0.2713&142&4051.1555&0.0564&150&4160.5846&0.0564&  150       \\
         &&            &          &&            &          &&            &          &&            \\
3866.5184&0.0564&150&3991.7376&0.2713&142&4059.9598&0.0564&150&4214.5814&0.0564&  150       \\
3866.5191&0.0576&148&3991.7385&0.1218&143&4059.9599&0.0576&148&4214.5900&0.0576&  148       \\
3866.5196&0.1719&146&3991.7397&0.2380&144&4059.9600&0.1719&146&4214.5955&0.1719&  146       \\
3866.5200&0.0830&145&3991.7401&0.0830&145&4059.9600&0.0830&145&4214.5996&0.0829&  145       \\
3866.5201&0.2380&144&3991.7418&0.1719&146&4059.9600&0.2380&144&4214.6008&0.2380&  144       \\
3866.5203&0.1218&143&3991.7441&0.0576&148&4059.9600&0.1218&143&4214.6036&0.1218&  143       \\
3866.5205&0.2713&142&3991.7476&0.0564&150&4059.9601&0.2713&142&4214.6059&0.2713&  142       \\
         &&            &          &&            &          &&            &          &&            \\
3869.0696&0.2713&142&4004.0083&0.0564&150&4061.0825&0.2713&142&4221.1369&0.0564&  150       \\
3869.0698&0.1218&143&4004.0091&0.0576&148&4061.0854&0.1218&143&4221.1383&0.0576&  148       \\
3869.0700&0.2380&144&4004.0096&0.1719&146&4061.0890&0.2380&144&4221.1393&0.1719&  146       \\
3869.0700&0.0830&145&4004.0100&0.0830&145&4061.0905&0.0830&145&4221.1399&0.0830&  145       \\
3869.0703&0.1719&146&4004.0101&0.2380&144&4061.0956&0.1719&146&4221.1401&0.2380&  144       \\
3869.0706&0.0576&148&4004.0103&0.1218&143&4061.1027&0.0576&148&4221.1406&0.1218&  143       \\
3869.0711&0.0564&150&4004.0105&0.2713&142&4061.1136&0.0564&150&4221.1410&0.2713&  142       \\
\enddata
\end{planotable}
}
\end{center}

% Second page: 44 l/page
% \documentstyle[apjpt4]{article}

% \begin{document}
% \makeatletter
% \def\jnl@aj{AJ}
% \ifx\revtex@jnl\jnl@aj\let\tablebreak=\nl\fi
% \makeatother

% \setcounter{page}{1}

\begin{center}
 
{
\small
% \bf

% \ptlandscape
% ALSO NEED TO DO   dvips $-$t landscape FILENAME   TO MAKE THE LANDSCAPE OPTION WORK PROPERLY

\begin{planotable}{lll|lll|lll|lll}
\tabletypesize{\tiny}
\tablewidth{0pt}
\tablenum{A3 continued}
\tablecaption{Neodymium isotope splitting}
\tablehead{
\colhead{$\lambda$ (\AA)} & \colhead{frac.} & \colhead{Label} &
\colhead{$\lambda$ (\AA)} & \colhead{frac.} & \colhead{Label} &
\colhead{$\lambda$ (\AA)} & \colhead{frac.} & \colhead{Label} &
\colhead{$\lambda$ (\AA)} & \colhead{frac.} & \colhead{Label} \\
}
\startdata
4232.3748&0.2713&142&4451.5767&0.1719&146&4477.4595&0.2713&142&4563.2230&0.0172&  146       \\
4232.3768&0.1218&143&4451.5851&0.0576&148&4477.4597&0.1219&143&4563.2267&0.0058&  148       \\
4232.3793&0.2380&144&4451.5980&0.0564&150&4477.4599&0.2380&144&4563.2324&0.0056&  150       \\
4232.3803&0.0830&145&          &         &            &4477.4600&0.0830&145&          &         &            \\
4232.3840&0.1719&146&4451.9873&0.2709&142&4477.4604&0.1719&146&4567.6067&0.2713&  142       \\
4232.3889&0.0576&148&4451.9883&0.1221&143&4477.4608&0.0575&148&4567.6080&0.1218&  143       \\
4232.3965&0.0564&150&4451.9896&0.2380&144&4477.4615&0.0564&150&4567.6096&0.2379&  144       \\
         &&            &4451.9902&0.0829&145&          &&            &4567.6102&0.0830&  145       \\
4256.8185&0.2713&142&4451.9921&0.1720&146&4501.8186&0.2713&142&4567.6125&0.1719&  146       \\
4256.8191&0.1218&143&4451.9946&0.0579&148&4501.8191&0.1218&143&4567.6156&0.0576&  148       \\
4256.8198&0.2380&144&4451.9986&0.0561&150&4501.8198&0.2380&144&4567.6204&0.0564&  150       \\
4256.8201&0.0829&145&          &&            &4501.8201&0.0830&145&          &&            \\
4256.8212&0.1719&146&4456.3910&0.2713&142&4501.8211&0.1719&146&4578.8885&0.2713&  142       \\
4256.8226&0.0576&148&4456.3945&0.1218&143&4501.8224&0.0576&148&4578.8891&0.1218&  143       \\
4256.8249&0.0564&150&4456.3988&0.2380&144&4501.8245&0.0564&150&4578.8898&0.2380&  144       \\
         &&            &4456.4006&0.0830&145&          &&            &4578.8901&0.0830&  145       \\
4358.1614&0.2713&142&4456.4068&0.1719&146&4506.5854&0.2713&142&4578.8911&0.1719&  146       \\
4358.1647&0.1218&143&4456.4153&0.0576&148&4506.5872&0.1219&143&4578.8925&0.0576&  148       \\
4358.1689&0.2380&144&4456.4284&0.0564&150&4506.5894&0.2380&144&4578.8946&0.0564&  150       \\
4358.1705&0.0830&145&          &&            &4506.5903&0.0829&145&          &&            \\
4358.1765&0.1719&146&4462.4194&0.2713&142&4506.5935&0.1722&146&4579.3181&0.2713&  142       \\
4358.1845&0.0576&148&4462.4196&0.1218&143&4506.5979&0.0577&148&4579.3189&0.1218&  143       \\
4358.1970&0.0564&150&4462.4199&0.2380&144&4506.6046&0.0561&150&4579.3198&0.2380&  144       \\
         &&            &4462.4200&0.0830&145&          &&            &4579.3201&0.0830&  145       \\
4368.6384&0.2713&142&4462.4205&0.1719&146&4516.3560&0.2713&142&4579.3214&0.1719&  146       \\
4368.6390&0.1218&143&4462.4210&0.0576&148&4516.3575&0.1218&143&4579.3231&0.0576&  148       \\
4368.6398&0.2380&144&4462.4219&0.0564&150&4516.3595&0.2380&144&4579.3258&0.0564&  150       \\
4368.6401&0.0830&145&          &&            &4516.3602&0.0830&145&          &&            \\
4368.6412&0.1719&146&4462.9801&0.2713&142&4516.3630&0.1719&146&4594.4474&0.2713&  142       \\
4368.6428&0.0576&148&4462.9839&0.1218&143&4516.3669&0.0576&148&4594.4484&0.1218&  143       \\
4368.6451&0.0564&150&4462.9887&0.2380&144&4516.3727&0.0564&150&4594.4497&0.2379&  144       \\
         &&            &4462.9906&0.0830&145&          &&            &4594.4502&0.0830&  145       \\
4446.3852&0.2713&142&4462.9975&0.1719&146&4542.6001&0.2713&142&4594.4519&0.1719&  146       \\
4446.3871&0.1218&143&4463.0068&0.0576&148&4542.6012&0.1218&143&4594.4544&0.0576&  148       \\
4446.3894&0.2380&144&4463.0212&0.0564&150&4542.6026&0.2380&144&4594.4581&0.0564&  150       \\
4446.3903&0.0830&145&          &&            &4542.6032&0.0830&145&          &&            \\
4446.3936&0.1719&146&4465.5976&0.2713&142&4542.6052&0.1719&146&4645.7662&0.2713&  142       \\
4446.3981&0.0576&148&4465.5985&0.1218&143&4542.6079&0.0576&148&4645.7677&0.1218&  143       \\
4446.4050&0.0564&150&4465.5997&0.2380&144&4542.6121&0.0564&150&4645.7695&0.2380&  144       \\
         &         &            &4465.6001&0.0830&145&          &         &            &4645.7702&0.0830&  145       \\
4451.5611&0.2713&142&4465.6018&0.1719&146&4563.2161&0.0271&142&4645.7729&0.1719&  146       \\
4451.5646&0.1218&143&4465.6040&0.0576&148&4563.2176&0.0122&143&4645.7765&0.0576&  148       \\
4451.5688&0.2380&144&4465.6075&0.0564&150&4563.2195&0.0238&144&4645.7820&0.0564&  150       \\
4451.5705&0.0830&145&          &         &            &4563.2202&0.0083&145&          &         &            \\
\enddata
\end{planotable}
}
\end{center}

\begin{center}
 
{
\small
% \bf

% \ptlandscape
% ALSO NEED TO DO   dvips $-$t landscape FILENAME   TO MAKE THE LANDSCAPE OPTION WORK PROPERLY

\begin{planotable}{lll|lll|lll|lll}
\tabletypesize{\tiny}
\tablewidth{0pt}
\tablenum{A4}
\tablecaption{Samarium isotope splitting}
\tablehead{
\colhead{$\lambda$ (\AA)} & \colhead{frac.} & \colhead{Label} &
\colhead{$\lambda$ (\AA)} & \colhead{frac.} & \colhead{Label} &
\colhead{$\lambda$ (\AA)} & \colhead{frac.} & \colhead{Label} &
\colhead{$\lambda$ (\AA)} & \colhead{frac.} & \colhead{Label} \\
}
\startdata
3922.3965&0.2270&154&4237.6577&0.1129&148&4452.7339&0.1130&148&4566.2170&0.0311&  144       \\
3922.3978&0.2670&152&4237.6582&0.1379&149&4452.7355&0.1500&147&          &         &            \\
3922.4004&0.0740&150&4237.6596&0.0740&150&4452.7406&0.0310&144&4577.6891&0.2271&  154       \\
3922.4017&0.1380&149&4237.6623&0.2670&152&          &         &            &4577.6894&0.2670&  152       \\
3922.4022&0.1130&148&4237.6636&0.2271&154&4458.5038&0.2270&154&4577.6901&0.0741&  150       \\
3922.4031&0.1500&147&          &         &            &4458.5098&0.2670&152&4577.6904&0.1379&  149       \\
3922.4059&0.0310&144&4244.6987&0.2270&154&4458.5219&0.0739&150&4577.6906&0.1130&  148       \\
         &&            &4244.6992&0.2670&152&4458.5279&0.1380&149&4577.6908&0.1500&  147       \\
3941.8700&0.2270&154&4244.7002&0.0740&150&4458.5301&0.1130&148&4577.6915&0.0309&  144       \\
3941.8700&0.2670&152&4244.7007&0.1380&149&4458.5343&0.1501&147&          &&            \\
3941.8700&0.0740&150&4244.7008&0.1130&148&4458.5476&0.0311&144&4584.8290&0.2270&  154       \\
3941.8700&0.1380&149&4244.7012&0.1500&147&          &&            &4584.8293&0.2670&  152       \\
3941.8700&0.1130&148&4244.7023&0.0309&144&4499.4765&0.0312&144&4584.8301&0.0740&  150       \\
3941.8700&0.1500&147&          &&            &4499.4782&0.1501&147&4584.8305&0.1380&  149       \\
3941.8700&0.0310&144&4318.9276&0.2270&154&4499.4787&0.1129&148&4584.8306&0.1130&  148       \\
         &&            &4318.9321&0.2670&152&4499.4790&0.1380&149&4584.8309&0.1500&  147       \\
3979.2000&0.2270&154&4318.9414&0.0740&150&4499.4798&0.0741&150&4584.8318&0.0309&  144       \\
3979.2000&0.2670&152&4318.9460&0.1380&149&4499.4813&0.2668&152&          &&            \\
3979.2000&0.0740&150&4318.9477&0.1130&148&4499.4820&0.2269&154&4593.5400&0.2269&  154       \\
3979.2000&0.1380&149&4318.9509&0.1500&147&          &&            &4593.5400&0.2670&  152       \\
3979.2000&0.1130&148&4318.9612&0.0310&144&4536.5100&0.2267&154&4593.5400&0.0741&  150       \\
3979.2000&0.1500&147&          &&            &4536.5100&0.2669&152&4593.5400&0.1381&  149       \\
3979.2000&0.0310&144&4345.8578&0.2271&154&4536.5100&0.0743&150&4593.5400&0.1130&  148       \\
         &&            &4345.8586&0.2669&152&4536.5100&0.1380&149&4593.5400&0.1499&  147       \\
4035.1001&0.2270&154&4345.8603&0.0741&150&4536.5100&0.1130&148&4593.5400&0.0310&  144       \\
4035.1038&0.2670&152&4345.8611&0.1381&149&4536.5100&0.1501&147&          &&            \\
4035.1111&0.0740&150&4345.8614&0.1131&148&4536.5100&0.0311&144&4595.2900&0.2269&  154       \\
4035.1148&0.1380&149&4345.8620&0.1498&147&          &&            &4595.2900&0.2670&  152       \\
4035.1161&0.1130&148&4345.8638&0.0309&144&4537.9276&0.2270&154&4595.2900&0.0740&  150       \\
4035.1187&0.1500&147&          &&            &4537.9358&0.2670&152&4595.2900&0.1380&  149       \\
4035.1268&0.0310&144&4424.3095&0.0310&144&4537.9526&0.0740&150&4595.2900&0.1130&  148       \\
         &&            &4424.3242&0.1500&147&4537.9608&0.1380&149&4595.2900&0.1499&  147       \\
4048.6069&0.2271&154&4424.3289&0.1130&148&4537.9639&0.1130&148&4595.2900&0.0311&  144       \\
4048.6117&0.2669&152&4424.3313&0.1380&149&4537.9697&0.1500&147&          &&            \\
4048.6215&0.0741&150&4424.3379&0.0740&150&4537.9881&0.0310&144&4642.2283&0.2270&  154       \\
4048.6264&0.1381&149&4424.3513&0.2670&152&          &&            &4642.2326&0.2670&  152       \\
4048.6282&0.1131&148&4424.3579&0.2270&154&4543.9449&0.2270&154&4642.2413&0.0740&  150       \\
4048.6316&0.1498&147&          &&            &4543.9468&0.2670&152&4642.2457&0.1380&  149       \\
4048.6424&0.0309&144&4433.8783&0.2270&154&4543.9506&0.0740&150&4642.2473&0.1130&  148       \\
         &&            &4433.8789&0.2670&152&4543.9525&0.1380&149&4642.2503&0.1500&  147       \\
4118.5484&0.2270&154&4433.8802&0.0740&150&4543.9532&0.1130&148&4642.2600&0.0310&  144       \\
4118.5490&0.2670&152&4433.8808&0.1380&149&4543.9545&0.1500&147&          &&            \\
4118.5502&0.0740&150&4433.8811&0.1130&148&4543.9587&0.0311&144&4674.5881&0.2270&  154       \\
4118.5508&0.1380&149&4433.8815&0.1500&147&          &&            &4674.5925&0.2670&  152       \\
4118.5510&0.1130&148&4433.8829&0.0310&144&4552.6578&0.0309&144&4674.6014&0.0740&  150       \\
4118.5514&0.1500&147&          &&            &4552.6589&0.1500&147&4674.6058&0.1380&  149       \\
4118.5527&0.0310&144&4434.2951&0.0310&144&4552.6592&0.1131&148&4674.6074&0.1130&  148       \\
         &&            &4434.3071&0.1500&147&4552.6594&0.1379&149&4674.6105&0.1500&  147       \\
4220.6553&0.2270&154&4434.3109&0.1130&148&4552.6599&0.0741&150&4674.6202&0.0310&  144       \\
4220.6570&0.2671&152&4434.3129&0.1380&149&4552.6608&0.2670&152&          &&            \\
4220.6605&0.0740&150&4434.3183&0.0740&150&4552.6613&0.2270&154&4687.1304&0.0311&  144       \\
4220.6623&0.1380&149&4434.3292&0.2670&152&          &         &            &4687.1543&0.1499&  147       \\
4220.6629&0.1130&148&4434.3346&0.2270&154&4566.2059&0.2269&154&4687.1619&0.1129&  148       \\
4220.6642&0.1500&147&          &         &            &4566.2074&0.2670&152&4687.1659&0.1380&  149       \\
4220.6680&0.0309&144&4452.7238&0.2270&154&4566.2105&0.0740&150&4687.1766&0.0740&  150       \\
         &         &            &4452.7261&0.2670&152&4566.2120&0.1380&149&4687.1984&0.2669&  152       \\
4237.6538&0.0310&144&4452.7307&0.0740&150&4566.2126&0.1131&148&4687.2091&0.2271&  154       \\
4237.6568&0.1500&147&4452.7330&0.1380&149&4566.2136&0.1501&147&          &         &            \\
\enddata
\end{planotable}
}
\end{center}
\begin{center}
 
{
\small
% \bf

% \ptlandscape
% ALSO NEED TO DO   dvips $-$t landscape FILENAME   TO MAKE THE LANDSCAPE OPTION WORK PROPERLY

\begin{planotable}{lll|lll|lll|lll}
\tabletypesize{\tiny}
\tablewidth{0pt}
\tablenum{A5}
\tablecaption{Europium hyperfine and isotope splitting}
\tablehead{
\colhead{$\lambda$ (\AA)} & \colhead{frac.} & \colhead{Label\tablenotemark{a}} &
\colhead{$\lambda$ (\AA)} & \colhead{frac.} & \colhead{Label} &
\colhead{$\lambda$ (\AA)} & \colhead{frac.} & \colhead{Label} &
\colhead{$\lambda$ (\AA)} & \colhead{frac.} & \colhead{Label} \\
}
\startdata
3819.5862&0.0371& 1.5-2.5 151&3907.1726&0.0271& 1.5-1.5 151&4129.7041&0.0429& 3.5-3.5 153&4435.5522&0.0006& 5.5-4.5 153\\
3819.6025&0.0476& 2.5-3.5 151&3907.1726&0.0148& 1.5-0.5 151&4129.7061&0.0146& 3.5-2.5 153&4435.5762&0.0984& 4.5-5.5 153\\
3819.6050&0.0079& 2.5-2.5 151&3907.1853&0.0053& 0.5-1.5 151&4129.7163&0.0153& 4.5-5.5 153&4435.5791&0.0191& 4.5-4.5 153\\
3819.6257&0.0613& 3.5-4.5 151&3907.1853&0.0185& 0.5-0.5 151&4129.7192&0.0099& 5.5-6.5 151&4435.5820&0.0014& 4.5-3.5 153\\
3819.6289&0.0124& 3.5-3.5 151&          &&            &4129.7192&0.0606& 4.5-4.5 153&4435.5903&0.0726& 3.5-4.5 151\\
3819.6313&0.0005& 3.5-2.5 151&3930.4368&0.0162& 5.5-4.5 151&4129.7217&0.0166& 4.5-3.5 153&4435.5967&0.0208& 3.5-3.5 151\\
3819.6538&0.0371& 1.5-2.5 153&3930.4409&0.1267& 5.5-5.5 151&4129.7271&0.0858& 5.5-5.5 151&4435.6016&0.0017& 3.5-2.5 151\\
3819.6560&0.0779& 4.5-5.5 151&3930.4810&0.0244& 4.5-3.5 151&4129.7334&0.0153& 5.5-4.5 151&4435.6016&0.0726& 3.5-4.5 153\\
3819.6597&0.0139& 4.5-4.5 151&3930.4841&0.0787& 4.5-4.5 151&4129.7344&0.0099& 5.5-6.5 153&4435.6040&0.0208& 3.5-3.5 153\\
3819.6611&0.0476& 2.5-3.5 153&3930.4885&0.0162& 4.5-5.5 151&4129.7378&0.0858& 5.5-5.5 153&4435.6064&0.0017& 3.5-2.5 153\\
3819.6621&0.0079& 2.5-2.5 153&3930.4944&0.0162& 5.5-4.5 153&4129.7407&0.0153& 5.5-4.5 153&4435.6211&0.0520& 2.5-3.5 153\\
3819.6628&0.0007& 4.5-3.5 151&3930.4963&0.1267& 5.5-5.5 153&4129.7598&0.1197& 6.5-6.5 153&4435.6235&0.0182& 2.5-2.5 153\\
3819.6714&0.0613& 3.5-4.5 153&3930.5139&0.0244& 4.5-3.5 153&4129.7632&0.0099& 6.5-5.5 153&4435.6250&0.0014& 2.5-1.5 153\\
3819.6729&0.0124& 3.5-3.5 153&3930.5154&0.0787& 4.5-4.5 153&4129.7764&0.1197& 6.5-6.5 151&4435.6257&0.0358& 1.5-2.5 153\\
3819.6738&0.0005& 3.5-2.5 153&3930.5171&0.0256& 3.5-2.5 151&4129.7842&0.0099& 6.5-5.5 151&4435.6357&0.0520& 2.5-3.5 151\\
3819.6848&0.0779& 4.5-5.5 153&3930.5173&0.0162& 4.5-5.5 153&          &&            &4435.6372&0.0120& 1.5-1.5 153\\
3819.6863&0.0139& 4.5-4.5 153&3930.5198&0.0455& 3.5-3.5 151&4204.9155&0.0238& 1.5-0.5 151&4435.6401&0.0182& 2.5-2.5 151\\
3819.6877&0.0007& 4.5-3.5 153&3930.5232&0.0244& 3.5-4.5 151&4204.9185&0.0119& 1.5-1.5 151&4435.6436&0.0014& 2.5-1.5 151\\
3819.6929&0.0979& 5.5-6.5 151&3930.5300&0.0256& 3.5-2.5 153&4204.9233&0.0013& 1.5-2.5 151&4435.6445&0.0238& 0.5-1.5 153\\
3819.6973&0.0126& 5.5-5.5 151&3930.5313&0.0455& 3.5-3.5 153&4204.9414&0.0357& 2.5-1.5 151&4435.6680&0.0358& 1.5-2.5 151\\
3819.7012&0.0006& 5.5-4.5 151&3930.5327&0.0244& 3.5-4.5 153&4204.9463&0.0182& 2.5-2.5 151&4435.6714&0.0120& 1.5-1.5 151\\
3819.7012&0.0979& 5.5-6.5 153&3930.5425&0.0212& 2.5-1.5 153&4204.9536&0.0017& 2.5-3.5 151&4435.6880&0.0238& 0.5-1.5 151\\
3819.7031&0.0126& 5.5-5.5 153&3930.5435&0.0245& 2.5-2.5 153&4204.9785&0.0520& 3.5-2.5 151&          &&            \\
3819.7048&0.0006& 5.5-4.5 153&3930.5444&0.0256& 2.5-3.5 153&4204.9854&0.0208& 3.5-3.5 151&4522.4878&0.0164& 5.5-4.5 151\\
3819.7205&0.1212& 6.5-7.5 153&3930.5454&0.0212& 2.5-1.5 151&4204.9946&0.0013& 3.5-4.5 151&4522.5010&0.1286& 5.5-5.5 151\\
3819.7227&0.0081& 6.5-6.5 153&3930.5471&0.0245& 2.5-2.5 151&4205.0166&0.0238& 1.5-0.5 153&4522.5513&0.0798& 4.5-4.5 151\\
3819.7246&0.0003& 6.5-5.5 153&3930.5498&0.0256& 2.5-3.5 151&4205.0181&0.0119& 1.5-1.5 153&4522.5410&0.0247& 4.5-3.5 151\\
3819.7366&0.1212& 6.5-7.5 151&3930.5515&0.0132& 1.5-0.5 153&4205.0205&0.0013& 1.5-2.5 153&4522.5620&0.0164& 5.5-4.5 153\\
3819.7417&0.0081& 6.5-6.5 151&3930.5520&0.0129& 1.5-1.5 153&4205.0264&0.0727& 4.5-3.5 151&4522.5679&0.1286& 5.5-5.5 153\\
3819.7463&0.0003& 6.5-5.5 151&3930.5530&0.0212& 1.5-2.5 153&4205.0283&0.0357& 2.5-1.5 153&4522.5845&0.0261& 3.5-2.5 151\\
         &&            &3930.5574&0.0107& 0.5-0.5 153&4205.0303&0.0182& 2.5-2.5 153&4522.5854&0.0247& 4.5-3.5 153\\
3907.0361&0.1428& 5.5-4.5 151&3930.5579&0.0132& 0.5-1.5 153&4205.0337&0.0017& 2.5-3.5 153&4522.5898&0.0798& 4.5-4.5 153\\
3907.0830&0.0220& 4.5-4.5 151&3930.5657&0.0132& 1.5-0.5 151&4205.0356&0.0192& 4.5-4.5 151&4522.5928&0.0464& 3.5-3.5 151\\
3907.0830&0.0970& 4.5-3.5 151&3930.5669&0.0129& 1.5-1.5 151&4205.0444&0.0520& 3.5-2.5 153&4522.5957&0.0164& 4.5-5.5 153\\
3907.0938&0.1428& 5.5-4.5 153&3930.5688&0.0212& 1.5-2.5 151&4205.0469&0.0007& 4.5-5.5 151&4522.6035&0.0247& 3.5-4.5 151\\
3907.1145&0.0220& 4.5-4.5 153&3930.5786&0.0107& 0.5-0.5 151&4205.0479&0.0208& 3.5-3.5 153&4522.6045&0.0261& 3.5-2.5 153\\
3907.1145&0.0970& 4.5-3.5 153&3930.5798&0.0132& 0.5-1.5 151&4205.0518&0.0013& 3.5-4.5 153&4522.6084&0.0464& 3.5-3.5 153\\
3907.1213&0.0017& 3.5-4.5 151&          &&            &4205.0659&0.0727& 4.5-3.5 153&4522.6128&0.0247& 3.5-4.5 153\\
3907.1213&0.0323& 3.5-3.5 151&4129.6021&0.0092& 1.5-2.5 151&4205.0698&0.0192& 4.5-4.5 153&4522.6191&0.0218& 2.5-1.5 151\\
3907.1213&0.0613& 3.5-2.5 151&4129.6050&0.0279& 1.5-1.5 151&4205.0747&0.0007& 4.5-5.5 153&4522.6201&0.0218& 2.5-1.5 153\\
3907.1316&0.0017& 3.5-4.5 153&4129.6196&0.0146& 2.5-3.5 151&4205.0859&0.0985& 5.5-4.5 151&4522.6226&0.0247& 2.5-2.5 153\\
3907.1316&0.0323& 3.5-3.5 153&4129.6240&0.0317& 2.5-2.5 151&4205.0923&0.0985& 5.5-4.5 153&4522.6250&0.0247& 2.5-2.5 151\\
3907.1316&0.0613& 3.5-2.5 153&4129.6270&0.0092& 2.5-1.5 151&4205.0972&0.0126& 5.5-5.5 151&4522.6260&0.0261& 2.5-3.5 153\\
3907.1448&0.0041& 2.5-3.5 153&4129.6455&0.0166& 3.5-4.5 151&4205.0972&0.0126& 5.5-5.5 153&4522.6313&0.0135& 1.5-0.5 153\\
3907.1448&0.0330& 2.5-2.5 153&4129.6509&0.0429& 3.5-3.5 151&4205.1235&0.1296& 6.5-5.5 153&4522.6328&0.0131& 1.5-1.5 153\\
3907.1448&0.0343& 2.5-1.5 153&4129.6548&0.0146& 3.5-2.5 151&4205.1562&0.1296& 6.5-5.5 151&4522.6333&0.0261& 2.5-3.5 151\\
3907.1514&0.0041& 2.5-3.5 151&4129.6782&0.0153& 4.5-5.5 151&          &&            &4522.6353&0.0218& 1.5-2.5 153\\
3907.1514&0.0330& 2.5-2.5 151&4129.6826&0.0092& 1.5-2.5 153&4435.4634&0.1296& 5.5-6.5 151&4522.6387&0.0106& 0.5-0.5 153\\
3907.1514&0.0343& 2.5-1.5 151&4129.6841&0.0279& 1.5-1.5 153&4435.4722&0.0126& 5.5-5.5 151&4522.6401&0.0135& 0.5-1.5 153\\
3907.1541&0.0057& 1.5-2.5 153&4129.6851&0.0606& 4.5-4.5 151&4435.4795&0.0006& 5.5-4.5 151&4522.6445&0.0135& 1.5-0.5 151\\
3907.1541&0.0271& 1.5-1.5 153&4129.6904&0.0166& 4.5-3.5 151&4435.5332&0.0984& 4.5-5.5 151&4522.6479&0.0131& 1.5-1.5 151\\
3907.1541&0.0148& 1.5-0.5 153&4129.6904&0.0146& 2.5-3.5 153&4435.5405&0.0191& 4.5-4.5 151&4522.6538&0.0218& 1.5-2.5 151\\
3907.1599&0.0053& 0.5-1.5 153&4129.6924&0.0317& 2.5-2.5 153&4435.5449&0.1296& 5.5-6.5 153&4522.6616&0.0106& 0.5-0.5 151\\
3907.1599&0.0185& 0.5-0.5 153&4129.6938&0.0092& 2.5-1.5 153&4435.5469&0.0014& 4.5-3.5 151&4522.6650&0.0135& 0.5-1.5 151\\
3907.1726&0.0057& 1.5-2.5 151&4129.7017&0.0166& 3.5-4.5 153&4435.5488&0.0126& 5.5-5.5 153&          &         &            \\
\enddata
\tablenotetext{a}{F$_{\rm lower}$-F$_{\rm upper}$ and isotope}
\end{planotable}
}
\end{center}
\begin{center}
 
{
% \small
% \bf

% \ptlandscape
% ALSO NEED TO DO   dvips $-$t landscape FILENAME   TO MAKE THE LANDSCAPE OPTION WORK PROPERLY

\begin{planotable}{lll}
\tabletypesize{\tiny}
\tablewidth{0pt}
\tablenum{A6}
\tablecaption{Lead hyperfine and isotope splitting}
\tablehead{
\colhead{$\lambda$ (\AA)} & \colhead{frac.} & \colhead{Label}
}
\startdata
3739.9230&0.2260& 207cog     \\
3739.9300&0.5230& 208        \\
3739.9420&0.2360& 206        \\
3739.9510&0.0150& 204        \\
         &&            \\
4057.7662&0.1357& 207c       \\
4057.8018&0.5230& 208        \\
4057.8149&0.2360& 206        \\
4057.8179&0.0151& 207b       \\
4057.8261&0.0151& 204        \\
4057.8386&0.0753& 207a       \\
\enddata
\end{planotable}
}
\end{center}

% END OF APPENDIX
%\end{document}